\documentclass{article}

\usepackage{PRIMEarxiv}

\usepackage[utf8]{inputenc} %
\usepackage[T1]{fontenc}    %
\usepackage{hyperref}       %
\usepackage{url}            %
\usepackage{booktabs}       %
\usepackage{amsfonts}       %
\usepackage{nicefrac}       %
\usepackage{microtype}      %
\usepackage{lipsum}
\usepackage{caption}
\usepackage{subcaption}
\usepackage[table,xcdraw]{xcolor}

\usepackage{wrapfig}
\usepackage{fancyhdr}       %
\usepackage{graphicx}       %
\usepackage{float}
\usepackage{amsmath}

\graphicspath{{figures/}}     %

\usepackage{placeins}
\usepackage{soul,color}
\usepackage{multirow}

\title{Crystal Transformer: Self-learning neural language model for Generative and Tinkering Design of Materials 
\thanks{\textit{\underline{Citation}}: 
\textbf{L.W...J.H. . Crystal transformer for generative and tinkering materials design. DOI:000000/11111.}} 
}

\author{
 Lai Wei\\
 Department of Computer Science and Engineering\\
  University of South Carolina\\
  Columbia, SC 29201 \\
  \And
 Qinyang Li, Yuqi Song\\
 Department of Computer Science and Engineering\\
  University of South Carolina\\
  Columbia, SC 29201 \\  
   \And
  Edirisuriya M. D. Siriwardane, Stanislav Stefanov\\
 Department of Computer Science and Engineering\\
  University of South Carolina\\
  Columbia, SC 29201 \\  
   \And
 Fanglin Chen\\
 Department of Mechanical Engineering\\
  University of South Carolina\\
  Columbia, SC 29201 \\    
   \And
 Jianjun Hu *\\
 Department of Computer Science and Engineering\\
  University of South Carolina\\
  Columbia, SC 29201 \\
  \texttt{jianjunh@cse.sc.edu} \\
}

\begin{document}
\maketitle

\begin{abstract}

Self-supervised neural language models have recently achieved unprecedented success, from natural language processing to learning the languages of biological sequences and organic molecules. These models have demonstrated superior performance in the generation, structure classification, and functional predictions for proteins and molecules with learned representations. However, most of the masking-based pre-trained language models are not designed for generative design, and their black-box nature makes it difficult to interpret their design logic. Here we propose BLMM Crystal Transformer, a neural network based probabilistic generative model for generative and tinkering design of inorganic materials. Our model is built on the blank filling language model for text generation and has demonstrated unique advantages in learning the "materials grammars" in terms of high-quality generation, interpretability, and data efficiency. It can generate chemically valid materials compositions with as high as 89.7\% charge neutrality and 84.8\% balanced electronegativity, which has more than 4 and 8 times higher enrichment compared to enhanced random sampling. The probabilistic generation steps allow it to recommend generation or tinkering actions with explanation, which captures known materials chemistry and makes it useful for materials doping. Our models can be trained with fewer than 40,000 materials formulas demonstrating their high data efficiency compared to other pre-trained protein or molecule models trained with millions of samples. We have applied our model to discover a set of new materials as validated using DFT calculations. Our work thus not only brings the unsupervised transformer language models based generative artificial intelligence to inorganic materials but also has the potential to guide the development of better generative design models in the domain of biology (proteins) and organic molecules. A user-friendly web app has been developed and can be accessed freely at \url{www.materialsatlas.org/blmtinker}.

\end{abstract}

\keywords{deep learning \and language models \and materials generator \and materials discovery \and blank filling}

\section{Introduction}

Discovery of novel functional materials such as high-capacity and safe electrodes and electrolytes for batteries or room-temperature superconductors has the potential to transform diverse industries \cite{saal2020machine,liu2021machine}. However, due to the sophisticated relationships of materials composition-structure-properties, centuries of rational design strategies have only covered an extremely limited chemical design space, among which the screening based approaches for discovering new materials are constrained by the limited scale and lack of diversity of known materials while the tinkering method is impeded by the incomplete understanding of the function related mechanisms and factors\cite{zunger2021understanding}. With the progress of generative machine learning, we recently showed that generative adversarial networks could be trained to generate chemically valid materials compositions \cite{dan2020generative}. However, the black-box nature of the deep neural network-based generator makes it difficult to interpret the black-box GAN models in terms of the chemical knowledge they learn and how they exploit the learned implicit knowledge for a generation. On the other hand, it is well known that materials tinkering or doping is one of the most widely used approaches to explore new materials \cite{zunger2021understanding} due to many constraints imposed on the possible options. During these processes, chemists or materials scientists usually resort to their intuition, chemical knowledge, and expertise to select substitution or doping elements and proportions to tune the properties of the material \cite{bustarret2006superconductivity,xiao2012synthesis,li2014bismuth,ye2019enhanced,wang2021doping} by considering a variety of factors such as compatibility of oxidation states, charge neutrality, coordination number, atomic radius, and other heuristic knowledge. 

Recently, Margraf et al. suggest that "materials grammars" can be defined based on expert knowledge to narrow down the design space in generative materials design \cite{margraf2021heterogeneous}. However, it is challenging for human to explicitly enumerate such chemical grammars considering so many chemical context-based dependencies among elements of stable compounds. To address this issue, a model with the capability of automatically distilling knowledge from data is highly desirable as shown in both language learning intelligent machines \cite{wei2021frequency} and AI game-players such as AlphaZero \cite{silver2018general}. Indeed, recently, pretrained self-supervised learning models such as BERT and GPT-3 have been proven to be effective at learning language grammars \cite{wei2021frequency} for text sentence generation \cite{rothe2020leveraging,li2021pretrained} and have achieved superior performance for many downstream tasks such as reading comprehension and question-answering. These language models been further transferred to the domain of proteins \cite{brandes2021proteinbert,song2021pretraining,brandesproteinbert} and organic molecules \cite{yu2021review,chen2021extracting,rong2020self,zhang2021mg,schwaller2019molecular}.
In 2019, Alley et. \cite{alley2019unified} showed that self-supervised protein language models are effective at learning protein representations for downstream tasks such as solubility prediction \cite{thumuluri2022netsolp,rives2021biological}. Brandes et al. proposed ProteinBert\cite{brandes2022proteinbert}, which showed strong performance in a variety of protein property prediction tasks such as protein structure and post-translational modifications prediction using the learned protein representation. 

Deep language models have also been used for the generation of protein sequences  \cite{madani2020progen,wu2020signal} and molecules \cite{bagal2021molgpt,rothchild2021c5t5}. In \cite{kim2021generative}, Kim et al. combined a transformer encoder with a conditional variational autoencoder (cVAE) to achieve high performance molecule generation. However, almost all existing language models for protein or molecule generation so far work mainly as a black-box without interpretable explanation of the learned grammars or rules and has difficulty to incorporate domain knowledge. A similar generative VAE was also proposed in \cite{dollar2021attention}. However, all these generative language models do not explicitly model the generative process and work more like black-box generators.

Despite the success of deep language models in protein and molecule generation, no studies have been reported that successfully applied deep language models to inorganic materials composition generation, possibly due to their extremely short formulas. At the same time, mask prediction based language models have been used to learn latent representations of elements, as shown in the Atom2vec method \cite{zhou2018learning}. Unsupervised word embedding learning has also been shown capable of capturing latent materials knowledge from materials science literature, which can be efficiently encoded as information-dense word embeddings (vector representations of words) without human labeling or supervision \cite{tshitoyan2019unsupervised}. These learned embeddings can capture complex materials science concepts such as the underlying structure of the periodic table and structure-property relationships in materials. However, their method is not intended for generating new materials and they do not model the dependency relationships of the elements within material compositions.

Inspired by the transformer-based language models with state-of-the-art performance on a range of natural language processing tasks and  structure and function prediction of protein and organic molecules, here we propose BLMM, a self-supervised language modeling approach \cite{shen2020blank,devlin2018bert,radford2019language,liu2019roberta} for generative and tinkering design of inorganic materials compositions. Our BLMM crystal transformer is based on a special self-supervised blank-filling language model (BLM) \cite{shen2020blank} which is trained with materials composition/formula data in the form of unlabeled expanded element symbol sequences sorted by the element electronegativity. These materials composition sequences use a small vocabulary of 118 or less elements, which is much larger than the 20 amino acid elements in protein language models but is much smaller than the vocabulary of natural texts. Unlike natural language texts, materials composition sequences have strong constraints among the elements due to the requirements to form chemically valid and structurally stable structures, which involve complex atomic interactions from ionic or covalent bonds and oxidation states of constituent elements. Effective generation models thus are required to learn complex local and long-range dependencies and contexts that the transformer neural network models excel at detecting and modeling. Our probabilistic blank filling model has an advantage over the heuristic or data mining element substitution models \cite{hautier2011data,sun2019map} as it can consider the chemical context within the formulas rather than only element property compatibility. Our extensive de novo materials composition generation and materials tinkering show that our BLM based materials generators learn chemical grammars and achieve interpretable generation due to their probabilistic predictions of the generation actions/steps.

\section{Results}
\label{sec:headings}

\subsection{Generative and tinkering materials design as a blank-filling process}

A typical ternary material composition can be represented as $A_xB_yC_z$ where A/B/C are elements and x/y/z are the number of atoms of corresponding elements. The same rule applies to compounds with different number of elements. If we only consider the cases where x/y/z are integers, we can expand the formula into $A_1 A_2...A_x B_1 B_2...B_y C_1 C_2..C_z$. For example, $SrTiO_3$ can be expanded to $Sr\; Ti\;  O\;  O\;  O$, which becomes a regular sequence similar to a natural text sequence of words or a sequence of amino acids or a \hl SMILES representation of a molecule.

\begin{table}[th]
\centering
\caption{Composition generation as a canvas rewriting process}
\label{tab:canvas}
\begin{tabular}{lll}
\hline
\multicolumn{3}{c}{ Canvas rewriting with 4 actions: (E, \_E, E\_, \_E\_)}                                            \\ \hline
\multicolumn{1}{l|}{Step t}           & Action  & operation  \\ \hline
\multicolumn{1}{l|}{0. \underline{\#1}}          & \_E\_ &Replace \#1 blank with \_Ti\_ \\ \hline
\multicolumn{1}{l|}{1. \underline{\#1} Ti \underline{\#2}}    & E & Replace \#1 blank with Sr     \\ \hline
\multicolumn{1}{l|}{2. Sr Ti \underline{\#1}}     & E\_ & Replace \#1 blank with O\_    \\ \hline
\multicolumn{1}{l|}{3. Sr Ti O \underline{\#1}}   & E\_ & Replace \#1 blank with O\_    \\ \hline
\multicolumn{1}{l|}{4. Sr Ti O O \underline{\#1}} & E &Replace \#1 blank with O      \\ \hline
\multicolumn{1}{l|}{5. Sr Ti O O O}   &               &                \\ \hline
\end{tabular}
\end{table}

Generating a chemical formula $SrTiO_3$ as represented by its expanded element sequence $Sr\; Ti\;  O\;  O\;  O$ from scratch can be done by the following canvas rewriting process (Table \ref{tab:canvas}): It starts with a starting canvas with a single blank \#1 $\_$ (non-terminal). For each blank, there are four possible canvas replacement/rewriting actions for each possible element out of 118 elements (or a subset): (1) action E:replace a blank with element E; (2) action \_E: replace a blank with element E and insert a new blank on its left side, allowing further element insertion; (3) action E\_: replace a blank with element E and insert a new blank on its right side, allowing further element insertion; (4) action \_E\_: replace blank with element E and insert new blanks on both sides. In Table\ref{tab:canvas}, step1 selects the action \_E\_ with element Ti, it generates a canvas with two blanks \_Ti\_. The model then selects action E, which just replaces left blank with element Sr. The next two steps all select action E\_, which replaces the blank with element O and insert a blank on the right. The final step just replaces the blank with another oxygen element. In addition to generation from scratch, the above canvas rewriting process can be naturally used for materials tinkering: we only need to mask some atoms in a known materials formula as blanks, and the blank filling process works the same way as de novo generation.

The key for the machine learning to learn the blankfilling generative model is how to learn the dependency of the rewriting process (blank-filling) over the preexisting contexts from the corpse of known inorganic materials composition sequences. The conditional developmental process of the canvas rewriting is similar to the growth of body plan of living organisms based on the cellular sense of growth factor or morphogen gradient \cite{schwank2010regulation} in local cellular context. The rewriting process has also been modelled in synthesis of circuits and dynamic systems \cite{fan2001bond} using genetic programming. Here we use a transformer based blank-filling language model to learn the context based material composition generation process \cite{shen2020blank}.

\subsection{Crystal Transformer: Blank filling language model for materials composition generation }

Our generative design model is based on the text filling blank language model (BLM) \cite{shen2020blank}, which is different from other popular deep language models such as BERT \cite{devlin2018bert} and XL-Net\cite{yang2019xlnet}, which usually mask and predict 15\% of tokens conditioned on the remaining text. This strategy is great for representation learning but may not be optimal for generation. The BLM model directly models the probabilistic dependency of words within the sentences and uses it to guide the sentence generation or blank-filling process, which makes it capable to fill blanks with partially specified text and achieve fine-grain control of generation locations while respecting the preceding and following contexts. It also has the capability of filling variable number of missing tokens.

\begin{figure}[ht]
  \centering
  \includegraphics[width=0.8\linewidth]{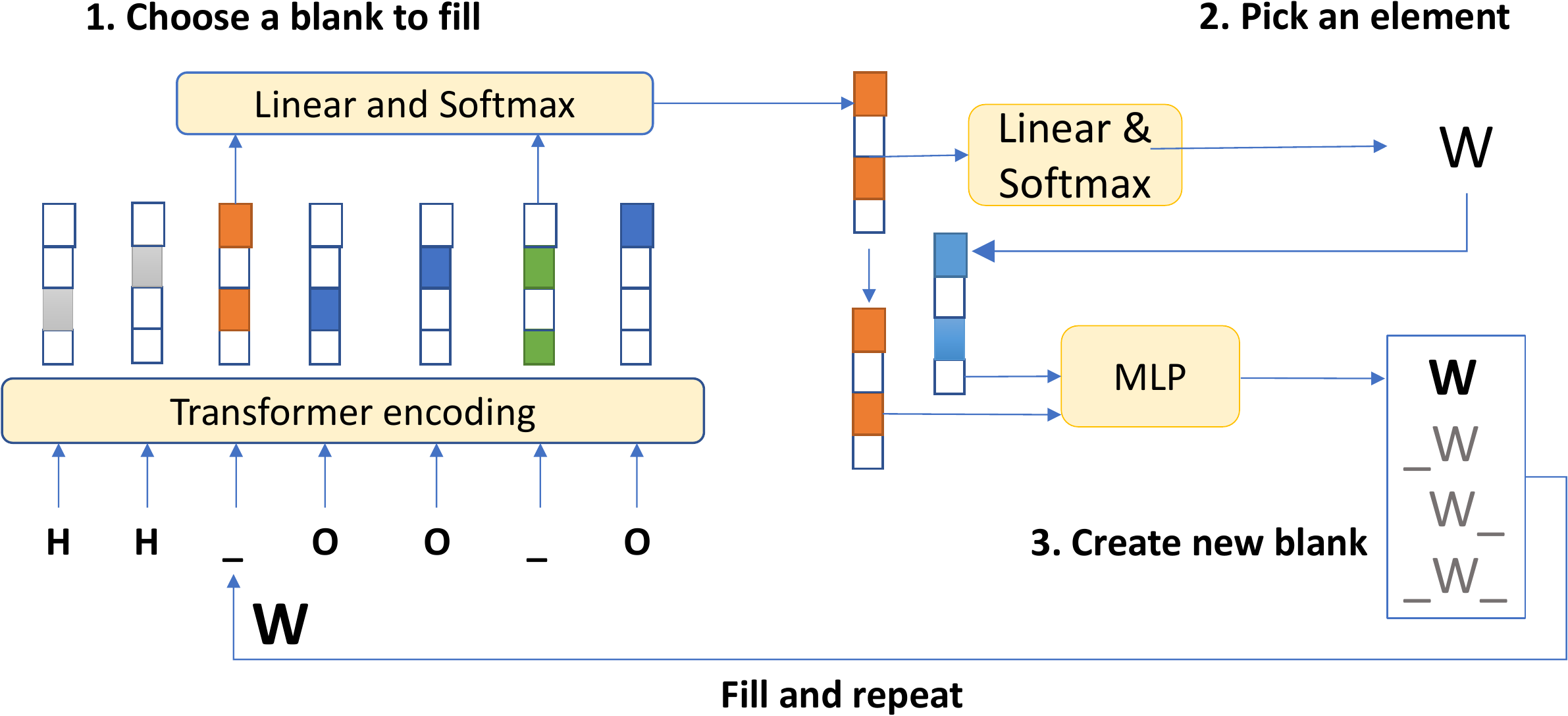}
  \caption{Neural network architecture of the blank filling language model for materials tinkering using H\textsubscript{2}WO\textsubscript{4} as an example.  }
  \label{fig:architecture}
\end{figure}

The architecture of our blank language model for materials (BLMM) is shown in Figure\ref{fig:architecture}. It consists of four main networks including the transformer network which encodes each of the tokens from the input, the expanded material formula with masked blanks, into position and semantic dependent embeddings. Then a blank selection network composed of linear and softmax layers will decide which blank to fill first. Next, the element selection network, also composed of linear and softmax layers, will pick an appropriate element to fill the selected blank. The embedding of the selected blank and the embedding of the picked element are then concatenated together and fed to the multi-layer Perceptron network to decide one out of four possible blank insertion options. Once the option is made, the canvas will be updated using the newly selected element and inserted blanks and the process will be repeated until all blanks have been filled. 

The training process works as follows: first randomly pick a training formula $x=\left(x_{1}, \cdots, x_{n}\right)$; then randomly sample $t$ between 0 and $n-1$ and sample an $n$-permutation $\sigma$, which indicates the generation order of the elements in the given formula. Now construct a canvas $c$ that keeps the first $t$ tokens $x_{\sigma_j}$ ($j=1,...,t)$ and collapse the remaining $n-t$ tokens as blanks. Next, get $n-t$ target actions $a_{j-t}$ for filling $x_{\sigma_j}$ ($j=t+1,...,n)$ into canvas. Then we compute the training loss by feeding the canvas $c$ into the neural networks and get the probabilities to pick the above determined actions. The loss is calculated as Eq. (1). More details can be found in \cite{shen2020blank}.

\begin{equation}
-\log (n !)-\frac{n}{n-t} \sum_{\sigma_{t+1}} \log p\left(a_{t}^{x, \sigma} \mid c_{t}^{x, \sigma} ; \theta\right)
\end{equation}
where the $\theta$ is the network weights; $c_{t}^{x, \sigma}$ is the $t$th canvas with the specified training formula $x$ and the selected generation order (permutation $\sigma$); $a_{t}^{x, \sigma}$ is the action at time $t$ which includes the blank selection, element picking, and blank option selection. 

In our BLMM model, we use the 118 elements plus 7 special tokens including $<PAD>, <UNK>, <FIRST>, <LAST>, <EOS>, <BLANK>, <BLANK\_0>$ as the vocabulary for training the BLMM models. If some elements are too infrequent, the model can remove those elements from the vocabulary.
The network models parameters are specified in the hyper-parameter part of the Method section.

\subsection{De novo generative design of materials composition }

\paragraph{Generation of hypothetical materials compositions:}
We prepare two sets of training datasets as described in Method to train different BLMM models for materials composition generation. The first set includes three datasets ICSD-mix, MP-mix, OQMD-mix, with selected compositions from ICSD, MaterialsProject, and OQMD databases respectively, all of which include samples that do not satisfy charge neutrality or balanced electronegativity. The second set of datasets includes ICSD-pure, MP-pure, OQMD-pure, which only includes selected formulas that are charge neutral with balanced electronegativity. For each of these datasets, we train a BLMM transformer model and use it to generate 100,000 hypothetical formulas.

\begin{figure}[ht!] 
    \begin{subfigure}[t]{0.33\textwidth}
        \includegraphics[width=\textwidth]{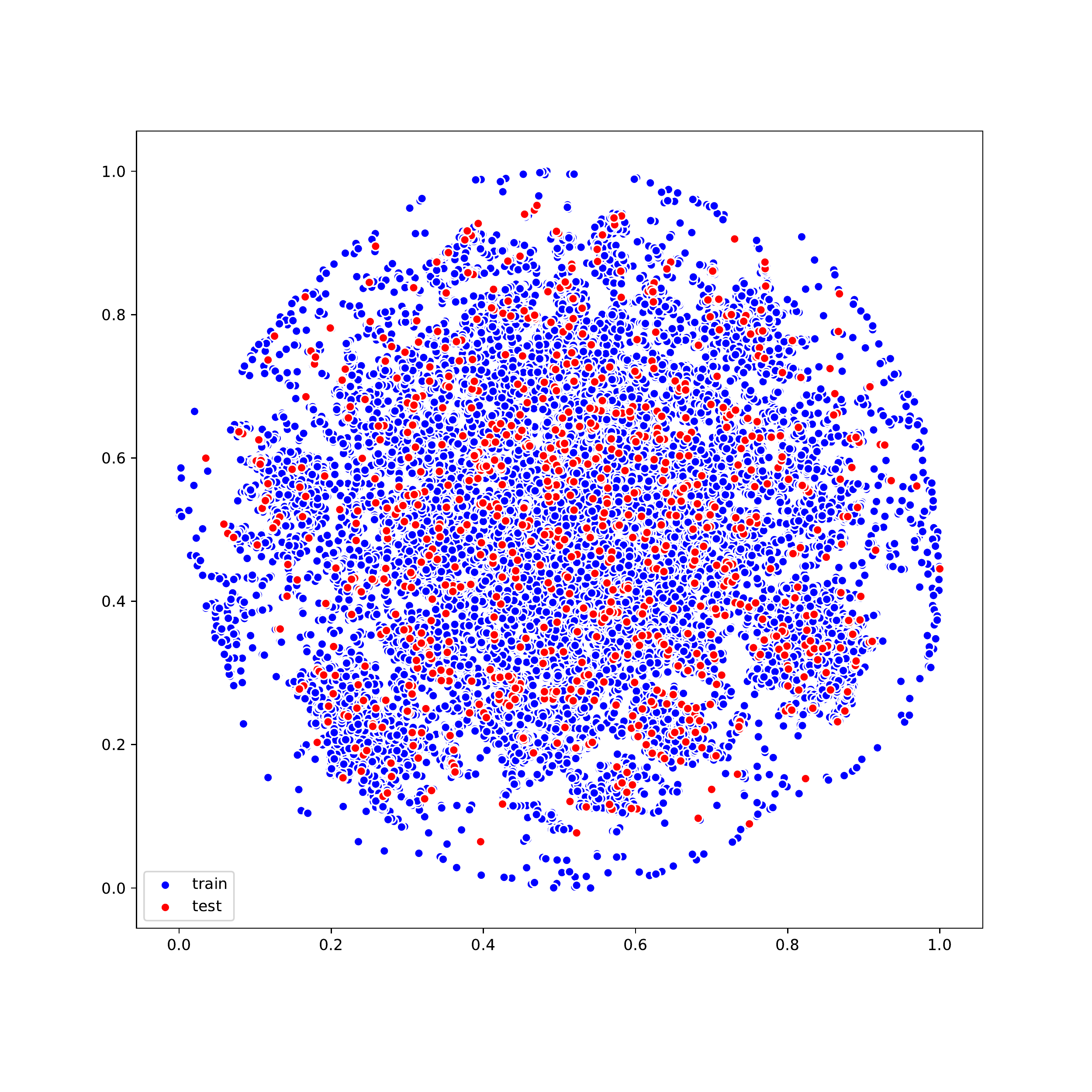}
        \caption{}
        \vspace{-3pt}
        \label{fig:GaB3N4_predict2}
    \end{subfigure}
    \begin{subfigure}[t]{0.33\textwidth}
        \includegraphics[width=0.99\textwidth]{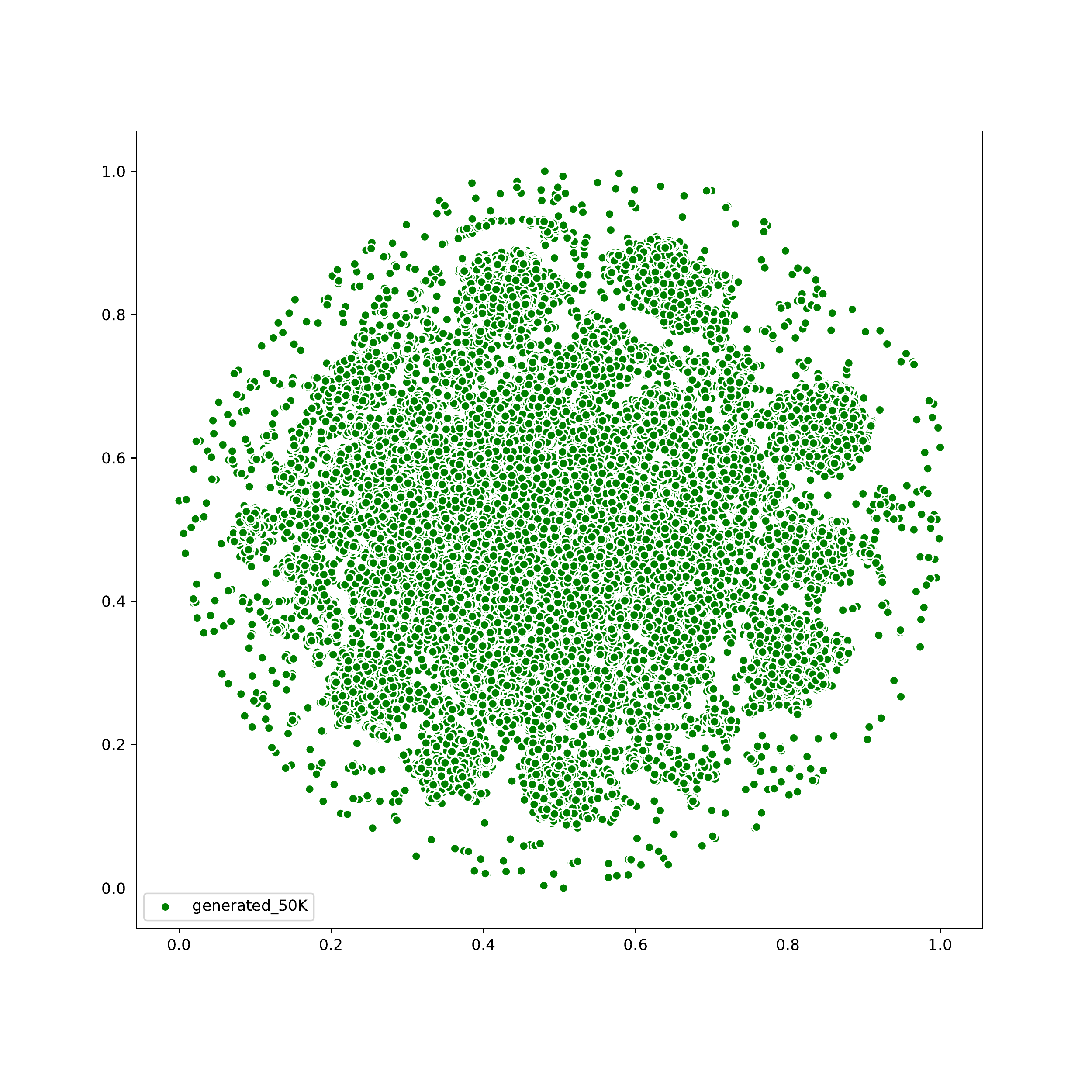}
        \caption{}
        \vspace{-3pt}
        \label{fig:GaB2N3_predict1}
    \end{subfigure} 
 \begin{subfigure}[t]{0.34\textwidth}
        \includegraphics[width=0.85\textwidth]{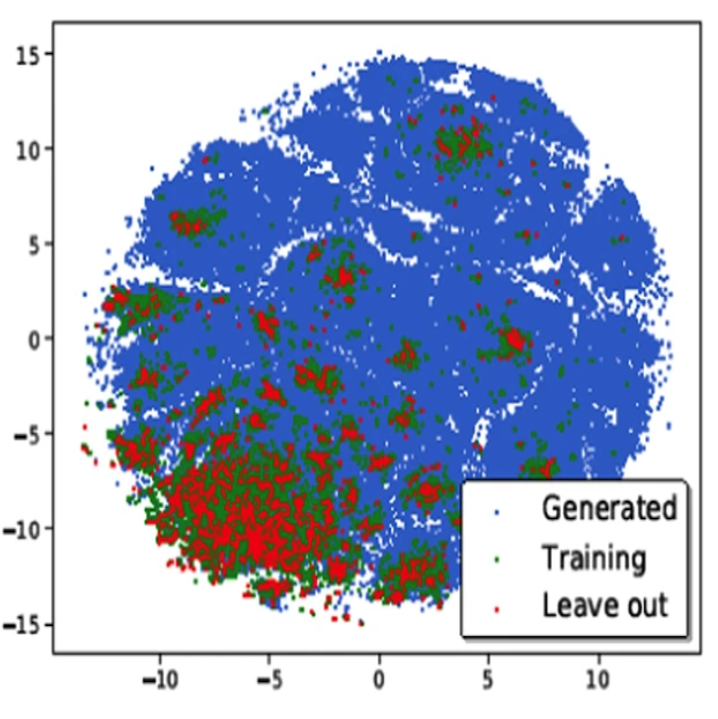}
        \caption{}
        \vspace{-3pt}
        \label{fig:GaB2N3_target}
    \end{subfigure}              

   \caption{Distribution of existing materials and  hypothetical materials generated by our BLMM-Pure-ICSD model. The distributions are generated by calculating the one-hot representation for the compositions and then use T-Sne to project them into 2-dimension space. (a) distribution of known training-testing formulas. (b) distribution of generated samples. (c) training-testing and generated samples of MATGAN \cite{dan2020generative}. }
  \label{fig:distribution}
\end{figure}

To evaluate whether our language BLMM models can learn the chemistry of inorganic materials (compositions) and use it to generate valid hypothetical formulas, we first check the distribution of the generated samples with respect to the training set and holdout test set of the Pure-ICSD dataset. We first represent each formula using the one-hot encoding as described in \cite{dan2020generative} and then map all the sample matrix representations into two-dimension space using the t-SNE algorithm. The results are shown in Figure\ref{fig:distribution}. First we find that the compositions  of existing materials in the ICSD dataset are not evenly distributed, but grouped into several clusters corresponding to materials families (Figure\ref{fig:distribution}(a)). We then find that the known materials (training and testing samples) are only a tiny portion of whole composition space and our generators can greatly expand the chemical composition design space (Figure\ref{fig:distribution}(b)). Compared to the distribution of generated samples (Figure\ref{fig:distribution}(c)) by MATGAN in \cite{dan2020generative}, our generated samples show much higher similarity with known materials.

\FloatBarrier

\subsection*{Evaluation of BLMM generation performance using validity, uniqueness, recovery rate, and novelty}

We evaluate the performance of our BLMM generators and compare with that of the baseline random formula generator using four evaluation criteria including validity, uniqueness, recovery rate, and novelty as described in the Method Section. 

\begin{figure}[ht!] 
    \begin{subfigure}[t]{0.49\textwidth}
        \includegraphics[width=\textwidth]{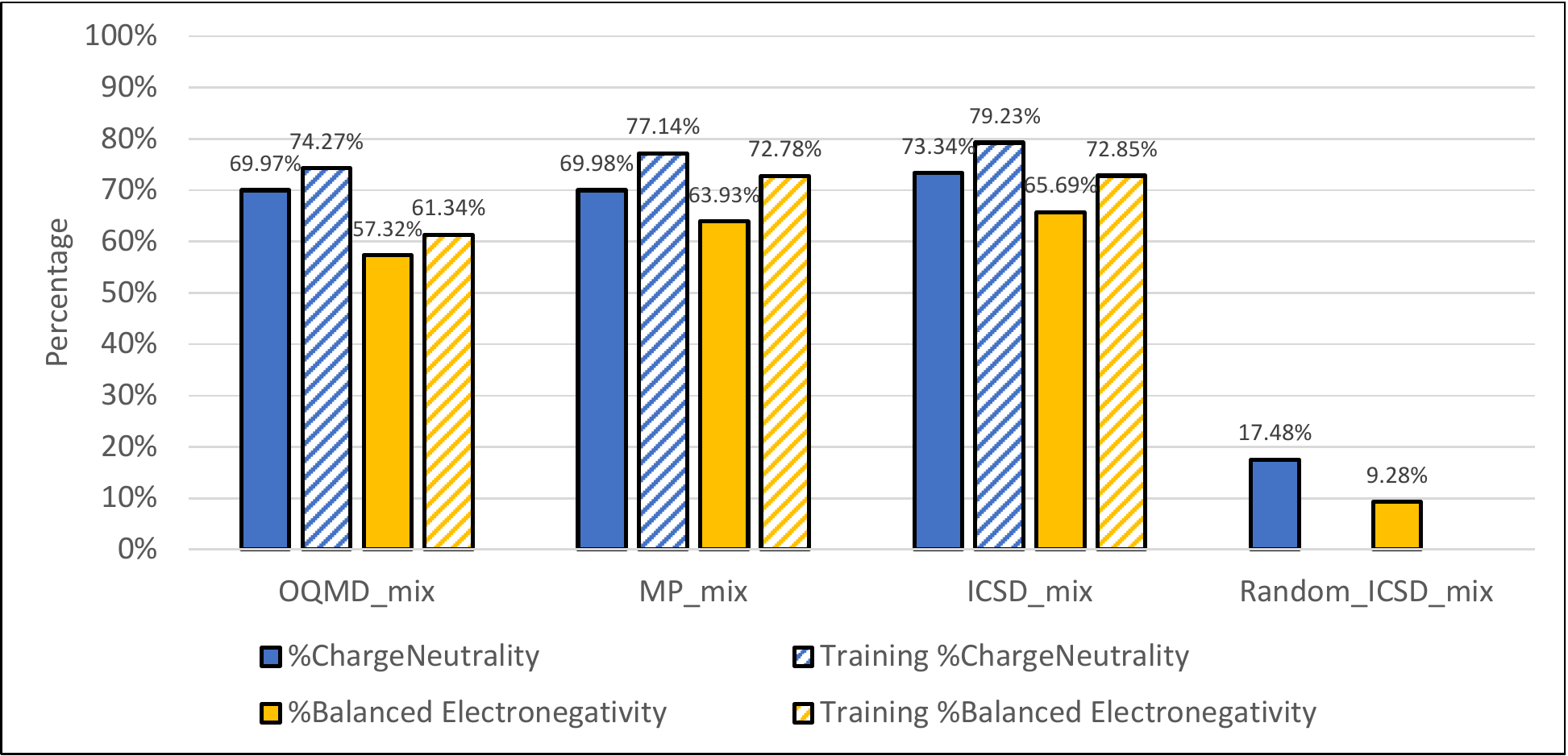}
        \caption{}
        \vspace{-3pt}
        \label{fig:mixChart}
    \end{subfigure}
    \begin{subfigure}[t]{0.49\textwidth}
        \includegraphics[width=\textwidth]{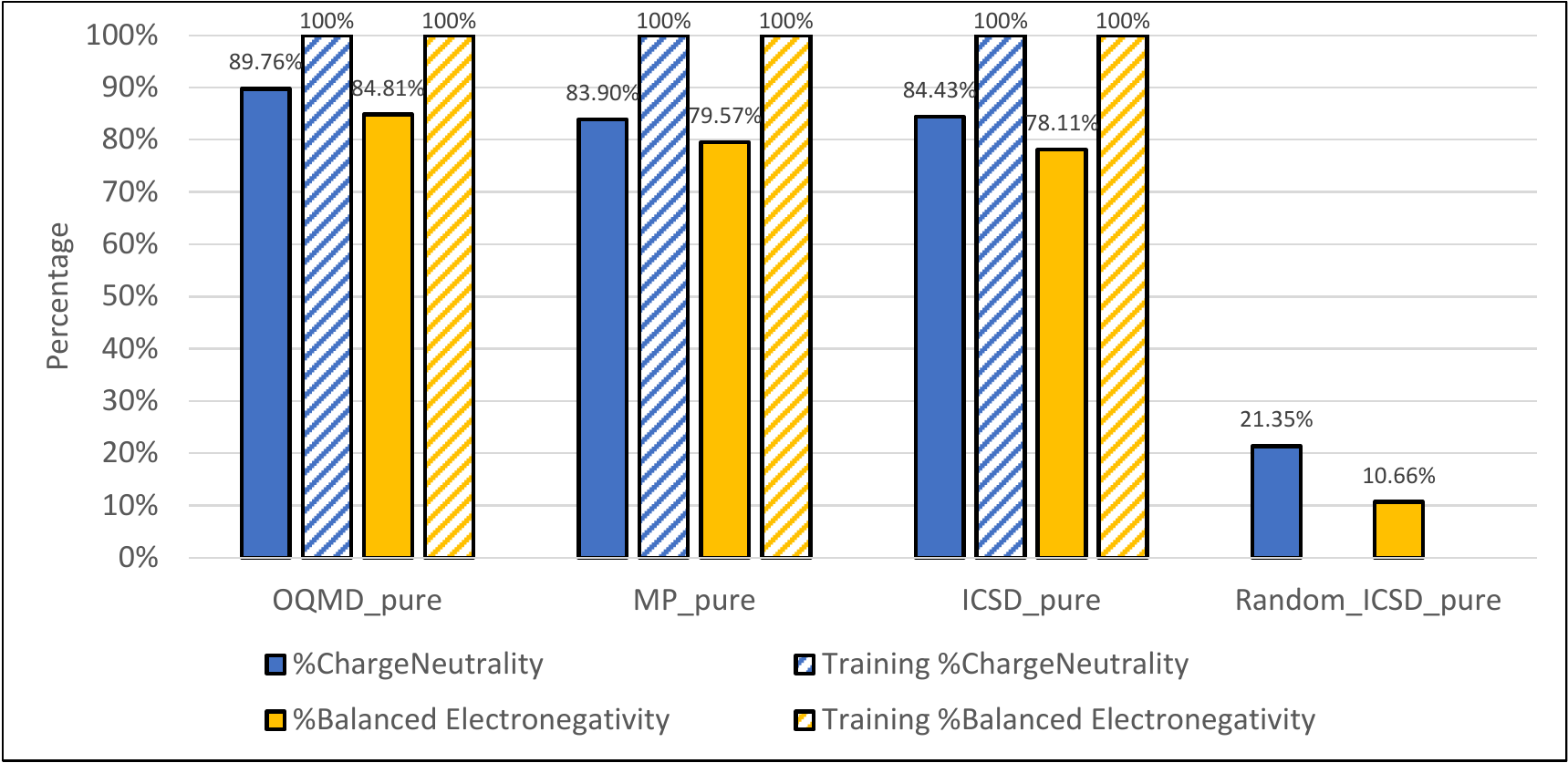}
        \caption{}
        \vspace{-3pt}
        \label{fig:pureChart}
    \end{subfigure} 
 \begin{subfigure}[t]{0.50\textwidth}
        \includegraphics[width=\textwidth]{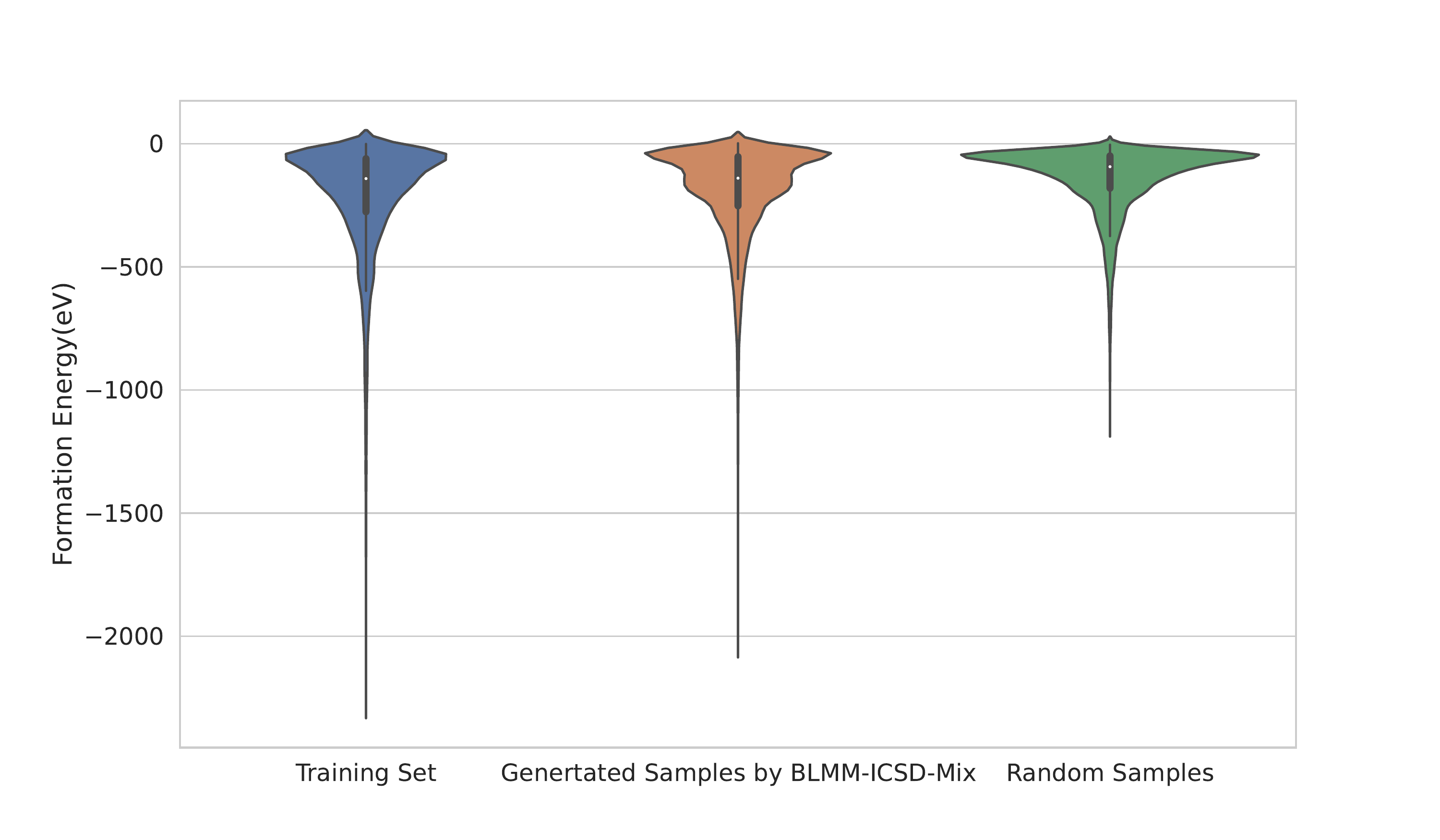}
        \caption{}
        \vspace{-3pt}
        \label{fig:GaB2N3_target}
    \end{subfigure}              
    \begin{subfigure}[t]{0.50\textwidth}
        \includegraphics[width=\textwidth]{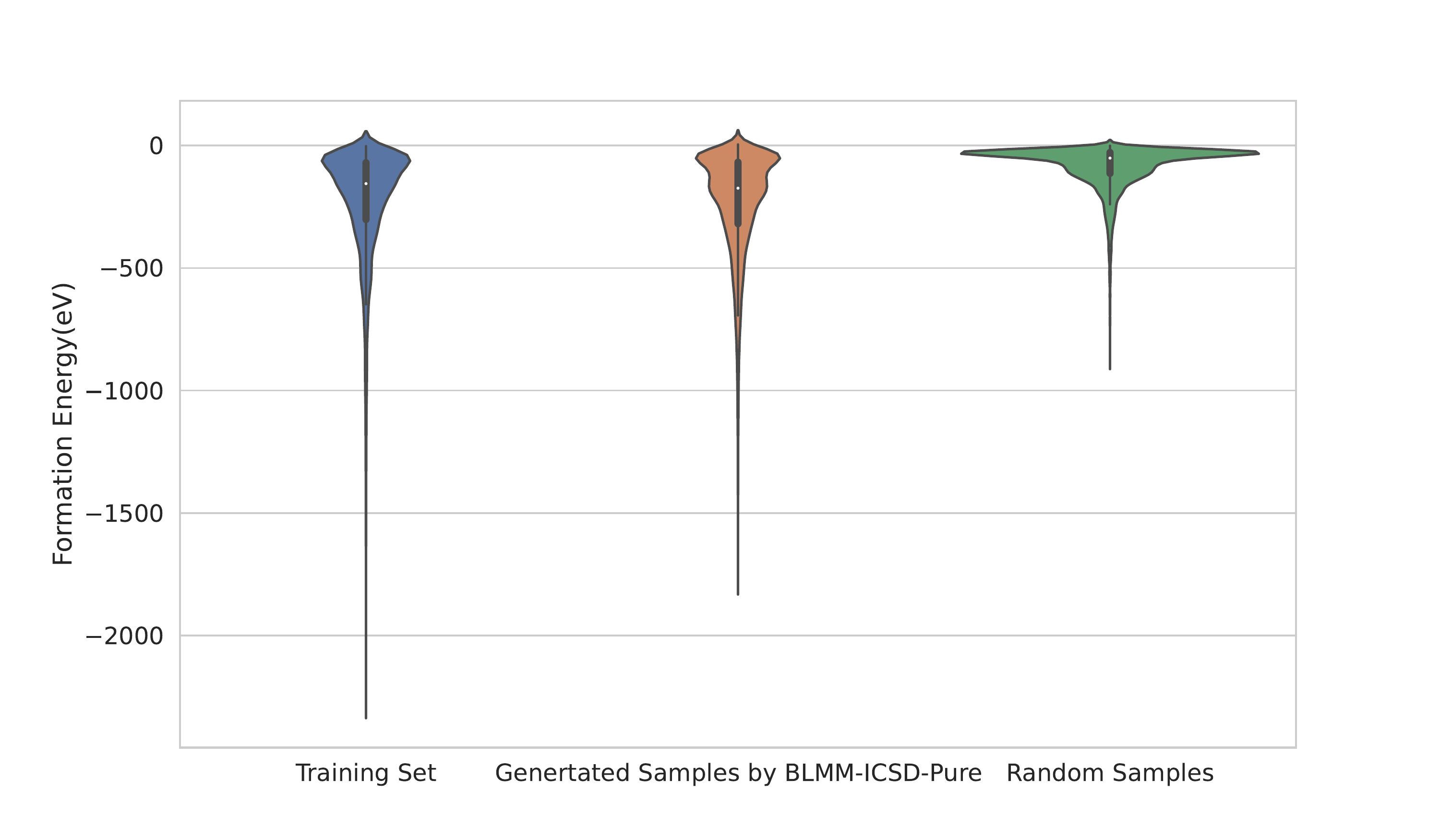}
        \caption{}
        \vspace{-3pt}
        \label{fig:GaB2N3_predict2}
    \end{subfigure}     

   \caption{Validity of BLMM materials composition generator. (a) The percentages of charge-neutral (CN) and electronegativity-balanced (EN) samples out of all generated samples by the BLMM models trained with mixed samples compared to those of the baseline random composition generator. (b) The percentage of charge-neutral (CN) and electronegativity-balanced (EN) samples out of all generated samples by the BLMM models trained with CN and EN samples compared to those of random composition generator. (c). the formation energy distribution of random samples, training set, and generated samples for BLMM-ICSD-mix models in (a). (d) the formation energy distribution of random samples, training set, and generated samples for BLMM-ICSD-pure models in (c). }
  \label{fig:validity}
\end{figure}

Figure\ref{fig:validity} (a) shows the composition generation performance of three BLMM models trained with three datasets OQMD-mix, MP-mix, and ICSD-mix. The OQMD-mix training dataset contains 345,022 materials compositions with 74.27\% samples satisfy charge neutrality (CN) and 61.34\% are electronegativity balanced (EN). Out of the 100,000 generated samples, we find that up to 69.97\% satisfy charge neutrality and 57.32\% meet balanced electronegativity, two of the major chemical validity requirements, indicating that our BLMM model has learned the chemical rules for assembling chemically valid compositions. The MP-mix training set has higher percentages in terms of charge neutrality and balanced electronegativity compared to OQMD-mix. It contains, however, only 84,664 samples. However, our BLMM models do not suffer from this significantly smaller dataset and still achieves high percentages of charge neutrality (69.98\%) and balanced electronegativity (63.93\%) for the 100,000 generated samples. While both OQMD-mix and MP-mix contain computationally derived materials (mainly via element substitutions), the ICSD-mix training set contains only 50,755 experimentally synthesized materials. Our BLMM model trained with this dataset shows even higher percentages of charge neutrality (73.34\%) and balanced electronegativity (65.69\%) for the 100,000 generated samples. As a comparison, we use the random composition generator (see Method) to generate 100,000 compositions using the anonymous formulas of the BLMM-ICSD-mix generated samples, which helps it to avoid the issue generating too random invalid formulas. Even with this lifting, the samples generated by the random generator only achieves 17.48\% of charge neutrality and 9.28\% in terms of balanced electronegativity. It is thus shown that our BLMM model achieves more than 4 and 6 times better performance in terms of generating chemically valid materials compositions compared to the random generator. 

We realize that the validity performance of our generators depends on the validity level of the training sets. To check if better training sets can improve the validity, we re-trained three models using the OQMD-pure, MP-pure, and ICSD-pure datasets, which all contain only samples that satisfy charge neutrality and balanced electronegativity. The results are shown in Figure\ref{fig:validity}(b). For the BLMM model trained with OQMD-pure, it now achieves charge neutrality of 89.76\% compared to 69.97\% of the model trained with OQMD-mix, a significant 19.78\% improvement despite its 40\% smaller dataset size (205,713 vs. 345,022), indicating the importance of data quality versus quantity for our BLMM model. For the BLMM models trained with MP-pure and ICSD-pure, similar significant validity performance are observed: the BLMM-MP-pure's charge neutrality percentage has been improved by 13.92\% and balanced electronegativity percentage by 15.64\%. For the BLMM-ICSD-pure model, these two validity performances have also been improved by 11.09\% and 12.42\%. In comparison, the lifted random generator only achieves charge-neutrality of 21.35\% and balanced electronegativity of 10.66\% for their 100,000 generated samples. 

We also compare the validity performance of our BLMM models with our previously developed MATGAN models that are based on generative adversarial network \cite{dan2020generative}. We find the GAN model trained with ICSD-mix achieves 80.3\% CN and 70.3\% EN compared to our 73.34\% CN and 65.69\% EN, with about 4-6\% advantage. Their GAN model trained with ICSD-pure achieves 92.1\% CN and 84.5\% EN compared to BLMM's 84.43\% and 78.11\%, also 6-7\% advantage. However, our hyper-parameter tuning experiments have shown that our BLMM models' performance can be further improved. The main interesting fact here is that as shown in Figure\ref{fig:distribution}, our BLMM models are complimentary to the MATGAN models: BLMM tend to generate hypothetical materials similar to the training samples, good for tinkering while MATGAN models are good for exploring new compositions. 

Another way to check the generation performance is to evaluate the stability of th generated compositions by predicting their formation energy. We first use the BLMM model trained with the ICSD-mix dataset to generate 100,000 samples and after filtering, 83,465 hypothetical samples remain. We then use the Roost based formation energy predictor (see Method) trained with all the Materials Project dataset to predict their formation energy. We also use the anonymous composition templates of these generated samples to create 83,465 random samples and predict their formation energy. The formation energy distributions of the ICSD-mix training set, the generated samples and random samples are shown in Figure\ref{fig:validity}(c). We find that the formation energy distribution of our BLMM-generated samples are much more similar to the training set compared to that of the random samples which have much more samples with formation energy closer to zero or above zero. We find the predicted formation energy of most of generated samples are lower than 0 eV, indicating their potential dynamic stability. Figure\ref{fig:validity}(d) shows the similar formation energy distributions for the training, generated, and random samples of the BLMM-ICSD-pure model, which contains 635,051 generated and random samples respectively.

\begin{figure}[ht!] 
    \begin{subfigure}[t]{0.5\textwidth}
        \includegraphics[height=0.7\textwidth]{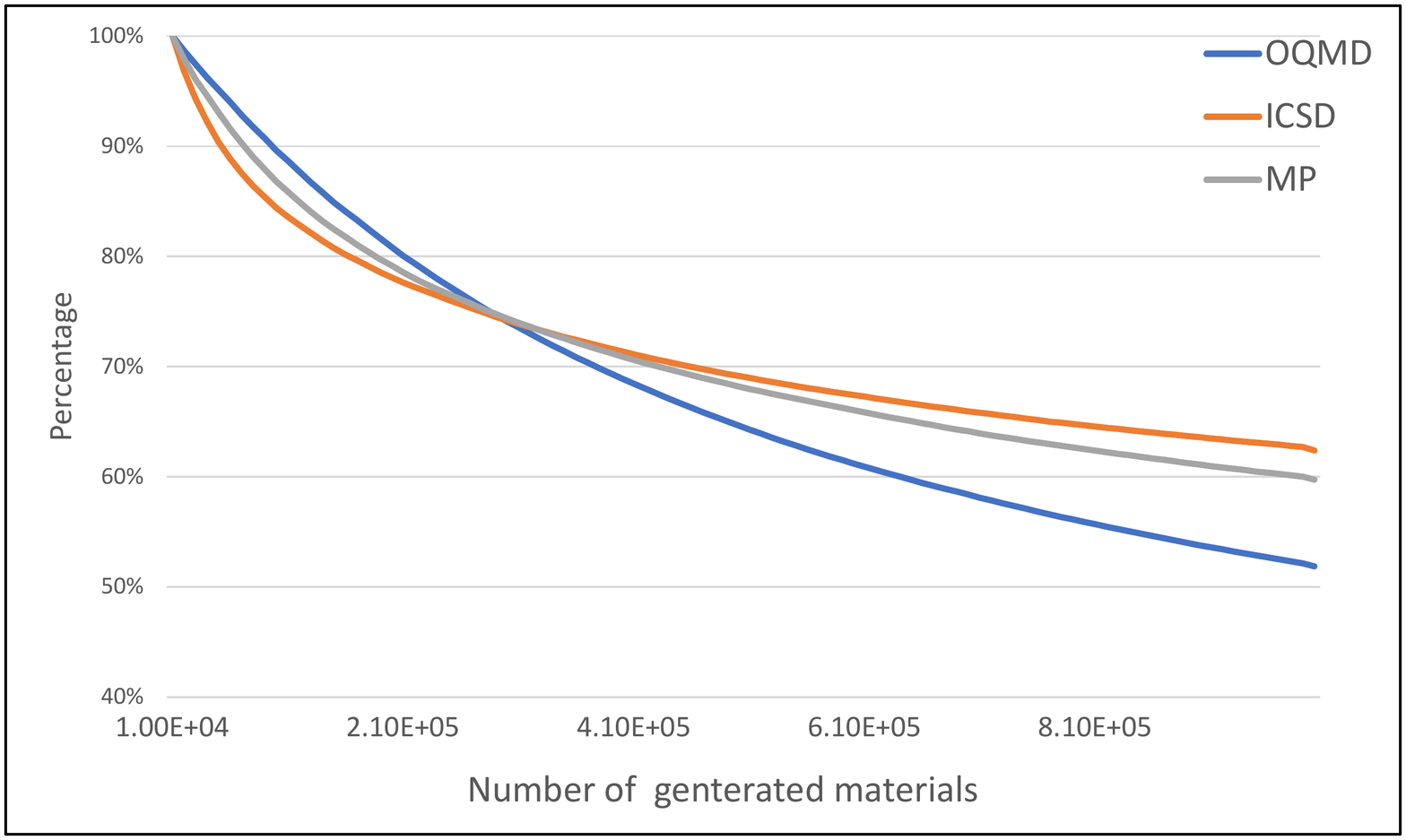}
        \caption{}
        \vspace{-3pt}
        \label{fig:GaB3N4_predict2}
    \end{subfigure}
    \begin{subfigure}[t]{0.5\textwidth}
        \includegraphics[height=0.67\textwidth]{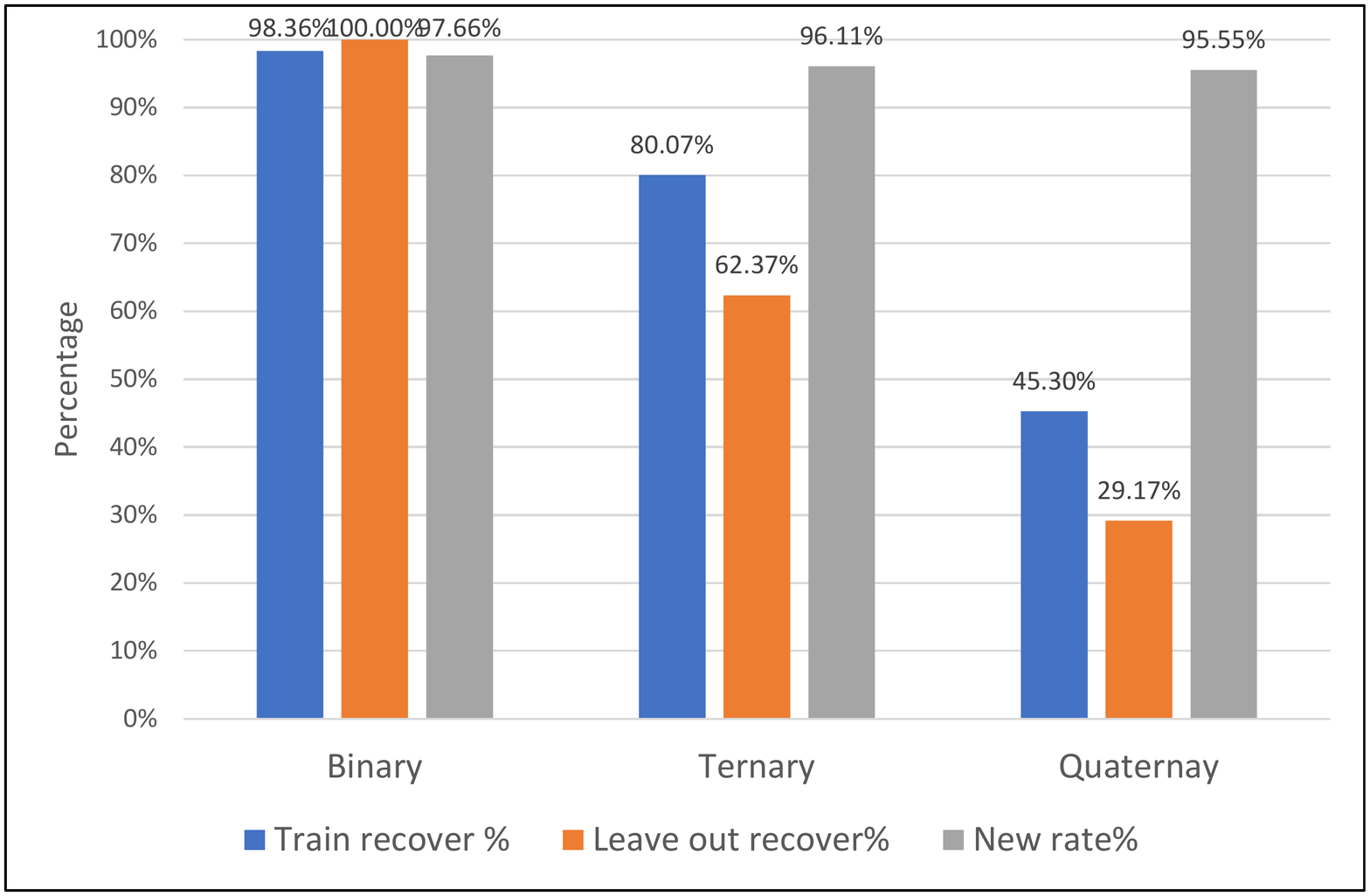}
        \caption{}
        \vspace{-3pt}
        \label{fig:GaB2N3_predict1}
    \end{subfigure} 
\caption{Uniqueness, recovery rate, and novelty of the BLMM generators trained with ICSD-pure, OQMD-pure, and MP-pure. (a) Comparison of the uniqueness curves of the generated samples. (b) Distribution of the recovery rates of the training and validation/testing samples as well as the novelty: the percentages of new generated hypothetical materials out of all generated samples}
  \label{fig:novelty}
\end{figure}

Another important performance measure of generators is the uniqueness, which calculates the percentage of unique samples out of all generated samples \cite{polykovskiy2020molecular}. Here for the three BLMM models trained with OQMD-pure, MP-pure, and ICSD-pure datasets, we calculate the uniqueness percentages at the end of every 10,000 generated samples up to 1 million. The results are shown in Figure\ref{fig:novelty}. First, we find that all three models have shown high uniqueness: even after generating one million samples, the uniqueness percentages remain above 50\%: OQMD (51.87\%), ICSD (62.38\%) and MP (59.73\%). The difference of these three models may be attributed to their different distributions of the training set. The OQMD dataset is mainly composed of ternary materials (>84.4\%) \cite{dan2020generative} so the BLMM-OQMD-pure model tends to generate ternary samples while the total number of chemically valid ternary compounds with integer ratios as estimated by SMACT (Semiconducting Materials from Analogy and Chemical Theory) to be around 200,000. So, it tends to generate more duplicate ternary samples. In contrast, the ratio of binary/ternary/quaternary is about 1:5.3:4.7 for the ICSD-pure dataset. For MP-pure dataset, it is about 1:7.3:6, which is much more balanced than previous two. It's interesting to find that before generating about 300,000 samples, the ICSD has the smallest uniqueness while OQMD has the highest one. The probable reason is that the OQMD dataset used here is much larger (216,540) compared to 63,703 of MP-pure and 39,431 of ICSD-pure), which cover more combinations of elements allowing it to generate diverse formulas in the beginning. However, after the inflection point at around 300,000 generated samples, the BLMM-OQMD model has visited most of the ternary formulas which it prefers to generate, causing it to fail to continue to generate new formulas, leading to lowest uniqueness. In contrast, the BLMM-ICSD is trained with more balanced binary, ternary and quarternary compounds, enabling it has the capability to generate diverse compositions even after 300,000 samplings. Compared to the MATGAN generators in \cite{dan2020generative}, we find that the uniqueness of our BLMM-OQMD is higher that that of GAN-OQMD while the BLMM-ICSD has the similar uniqueness of 75\% as their GAN-ICSD. However, their GAN-ICSD has much higher uniqueness of around 87\% than the 75\% of our BLMM-ICSD. This is likely due to that our BLMM is probabilistic model that uses neural networks to explicitly learn the context dependency among the elements within the compositions, which makes it tend to more closely approximate the elemental combinations. Instead, the GAN model used in MATGAN implicitly learns to approximate the distributions of training set as determined by the discriminator model, which makes them have less constraints that allow them to explore the chemical design space more freely.

We also check our BLMM models' capability for generating new materials compositions. We use the BLMM model trained with ICSD-pure to generate 1 million compositions and obtain 784,829 binary/ternary/quarternary compositions, which are used to calculate the recovery rate and novelty. First, we check whether our BLMM models can learns the chemical composition rules of the training set by calculating the training set recovery rates.  The results of our BLMM model are shown in Figure \ref{fig:novelty}(b) blue bars. Our model has recovered as much as 98.36\% of the 2,624 binary compounds in the training set due to the limited combinations of binary compositions. The model also recovers 80.07\% ternary compounds and 45.3\% quarternary compounds in the training set. These recovery rates are all significantly higher than those of our previous MATGAN models in \cite{dan2020generative} including 78.1\% for binary, 30.4\% for ternary, and 3.3\% for quarternary compounds. This much higher recovery rates of the training samples indicates our BLMM model's capability to learn known chemical composition distribution.

The recovery rates of our BLMM model on the hold-out test samples are shown as orange bars in Figure\ref{fig:novelty}. Our model achieves recovery percentages of 100\% of 59 binary materials, 63.37\% of 372 ternary materials, and 29.17\% of 360 quarternary ones. The reason that the quarternary compounds have much lower recovery rate is the quarternary design space is much higher \cite{davies2016computational}. Since these holdout ICSD samples are all experimentally synthesized materials which are not contained in the training set, the high holdout recovery rates directly demonstrate our model's capability to generate chemically valid real materials. Considering the huge quarternary compound space (~$4.1\times 10^{12}$), the 29.17\% recovery rate of our BLMM model to find the 105 samples out of 360 samples within only 1 million generated samples is like a feat comparable to finding a needle in the haystack. In contrast, the holdout recovery rates of MATGAN for binary, ternary and quarternary compounds are only 82.7\%, 31.2\%, and 5.2\%. Our BLMM model's holdout recovery rate for quarternary samples is 5.6 times of MATGAN. 

We also check the novelty of our BLMM model, which measures the percentage of the generated samples that are not within the known ICSD dataset. Our model achieves 97.66\%, 96.11\%, and 95.55\% for binary, ternary and quarternary compounds respectively, indicating its strong capability to explore new materials.

\begin{figure}[ht]
  \centering
  \includegraphics[width=0.6\linewidth]{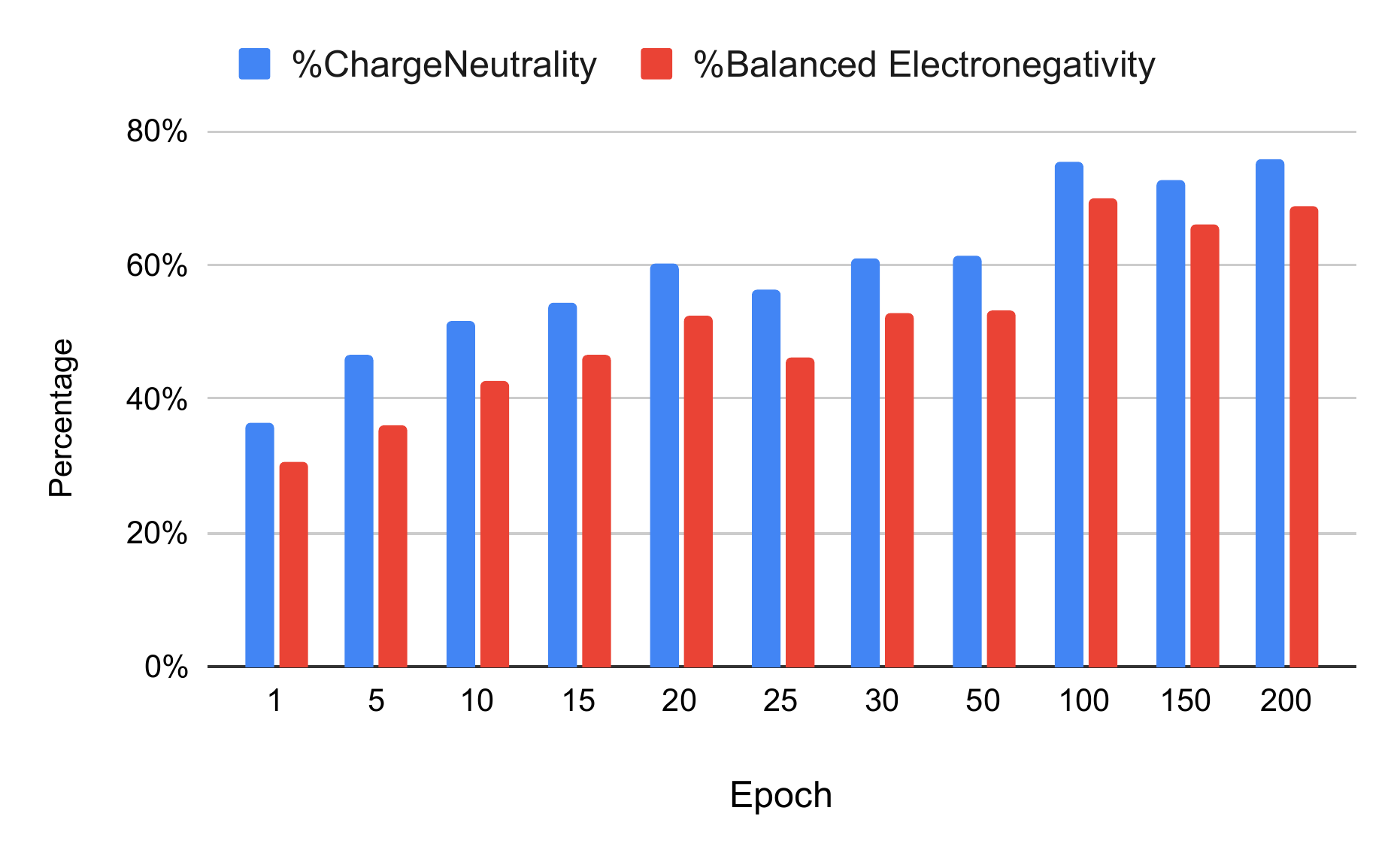}
  \caption{Increasing percentages of charge-neutral and electronegativity balanced samples generated by the models saved over the training process. In the beginning, few samples satisfy these two chemical validity rules. As the training goes on, the models gradually gain the capability to generate chemically valid materials compositions.}
  \label{fig:emergingorder}
\end{figure}

\FloatBarrier
\paragraph{Process of BLMM's learning of chemical rules:}  To illustrate the chemical order/rules emerge during the training process of our BLMM model, we save the intermediate models at the end of 1/5/10/15/20/25/30/50/100/150/200 epochs of training using the MP-mix dataset. We then generate 10,000 samples for each of these models and calculate the percentages of charge-neutral samples and electronegativity balanced samples. The results are shown in Figure\ref{fig:emergingorder}. We find that in the beginning, only a low percentage of the generated samples satisfy these two basic chemical rules: less than 20\% when the models are trained with less than 10 epochs. However, when the training epochs surpass 25 epochs, the percentage of charge neutral samples has already reached more than 50\% while the percentage of balanced electronegativity is slightly lower. When the training epochs reach 200, the \%chargeNeutrality has already reaches almost 70\% and \%balanced electronegativity reaches 64\%. 

To get intuitive understanding of orders shown in the generated samples, Table\ref{tab:generatedsamples} shows the typical generated oxide samples for the models saved at the end of the different epochs. At the epoch 1, the elements of the composition are almost randomly ordered. When the models are trained with 5 to 50 epochs, the formulas already include both anions are cations for creating charge-neutral compositions. However, the elements of the same type are still not completely ordered as the training samples. Even though formally, the appearance order of the elements in the formula does not make chemical difference, it is a violation of the orders within the training samples. When the training epochs reach 50, we find almost all the generated samples have the correct the element order: elements of the same type are shown together forming different clusters and the oxygen appears at the end of the formulas just as the training samples. This indicates that our BLMM model has learned the implicit chemical rules/order from our training set. Supplementary Table1 shows more generated samples by the models saved at epochs ranging from 1 to 200.

\begin{table}[th]
\centering
\caption{Emergence of orders within the generated samples over training epochs. In the beginning, the sequence of element symbols are mostly random. As training process goes on, the elements in the generated samples are more ordered by their electronegativity as the training samples show.}
\label{tab:generatedsamples}
\begin{tabular}{|r|l|l|}
\hline
\multicolumn{1}{|l|}{Epoch} & Samples                                 &Formulas  \\ \hline
1                           & \cellcolor[HTML]{FFFFFF}F I As O Rb O O K F F &KRbAsI(OF)$_3$  \\ \hline
5                           & \cellcolor[HTML]{FFFFFF}O O Na S Rh O O O O O &NaRhSO$_7$  \\ \hline
10                          & \cellcolor[HTML]{FFFFFF}Cd Mo O Mo O Mo O O &CdMo$_3$O$_4$  \\ \hline
15                          & \cellcolor[HTML]{FFFFFF}Cr O F O O F &CrO$_3$F$_2$  \\ \hline
20                          & \cellcolor[HTML]{FFFFFF}H H N O Cl O O &H$_2$NClO$_3$  \\ \hline
25                          & \cellcolor[HTML]{FFFFFF}Ta Ta Se N Se O O O O &Ta$_2$Se$_2$NO$_4$  \\ \hline
30                          & \cellcolor[HTML]{FFFFFF}Ba Ba Ge F O F F F O &Ba$_2$Ge(OF$_2$)$_2$  \\ \hline
50                          & \cellcolor[HTML]{FFFFFF}Ba Ba Fe Fe O Cl O O &Ba$_2$Fe$_2$ClO$_3$  \\ \hline
100                         & \cellcolor[HTML]{FFFFFF}Sm Sm Sm Ni O O O O O O &Sm$_3$NiO$_6$  \\ \hline
150                         & \cellcolor[HTML]{FFFFFF}Li Li Ti Zn Zn O O O O O &Li$_2$TiZn$_2$O$_5$  \\ \hline
200                         & \cellcolor[HTML]{FFFFFF}Sr Bi Br O O O O O &SrBiBrO$_5$  \\ \hline
\end{tabular}
\end{table}

\FloatBarrier

\subsection{Tinkering design of materials using blank filling transformer networks}

\paragraph{Design of new materials using BLMM}
One of the major advantages of BLMM language model based composition generator compared to GAN based models \cite{dan2020generative} is that it allows conditional composition generation starting with the templates from known crystal materials, by which we can specify some elements a prior, and then the model will fill the remaining blanks. This can be very useful for exploratory materials search. To demonstrate this capability, we start with Perovskite SrTiO\textsubscript{3}. We mask Ti from the expanded formula sequence Sr \_ O O O, and feed it to our BLMM model trained with the MP-mix dataset, the model suggests a list of possible filling elements as shown in Table \ref{tab:doping} column 1 (only top 20 suggestions are shown here). The element with the highest ranking is Ti with a probability score of 0.051. The other suggested elements include Ru, Si, Ge, C, Sn, Mn, Ir, Pt, Cr, Mo, which all lead to valid material entries included in the Materials Project database. We then mask the Sr element from SrTiO\textsubscript{3} and feed it to the network, the model  suggests the Sr as the 3rd best candidate. The other suggestions include Ba, Ca, Mg, Cs, Zr
 which all form valid entries in Materials Project database. Three of the suggestions including Li, Rb,Ta are not in the database. We check the charge neutrality and electronegativity balance for these three hypothetical compositions LiTiO\textsubscript{3}, RbTiO\textsubscript{3}, TaTiO\textsubscript{3} using the Smact package as implemented at MaterialsAtlas.org website \cite{hu2021materialsatlas}. They all satisfy these two chemical rules, indicating their potential to be valid crystals.

We further mask the Fe element from the lithium ion battery cathode material LiFePO\textsubscript{4} and feed it to our model, which suggests Mn and Co with probabilities of 0.206 and 0.15, leading to LiMnPO\textsubscript{4} and LiCoPO\textsubscript{4} , two candidate cathode materials under study\cite{nitta2015li}. We find the probabilistic BLMM model is much more flexible compared to the GAN model for composition generation \cite{dan2020generative}. For example, our model can be used to find doping elements for tuning crystal properties. For example, we mask the Mn element in the Li\textsubscript{2}MnO\textsubscript{3}, a well-known ionic conductor and feed it to the network model, which suggests Co, Cr, Ni, Ti, Fe, all have been used as doping elements in experimental studies \cite{singh2012electrochemical,fu2014electrochemical}. 
To further prove our model can discover new materials, we mask the Ga element in a known material Sr\textsubscript{3}GaN\textsubscript{3}, which is not contained in the training set for our model. We feed the masked sequence Sr Sr Sr \_ O O O to our model, which not only identifies the masked element Ga, but also suggests Cr as a substitution element, which leads to the rediscovery of an important electrode material Sr\textsubscript{3}CrN\textsubscript{3} as studied in \cite{chanhom2019sr3crn3} and another known crystal Sr\textsubscript{3}FeN\textsubscript{3} \cite{kikkawa1996transition}.

\begin{table}[th]
\centering
\caption{BLMM for element substitution and materials doping}
\label{tab:doping}

\begin{tabular}{|
>{\columncolor[HTML]{FFFFFF}}r |
>{\columncolor[HTML]{FFFFFF}}r |
>{\columncolor[HTML]{FFFFFF}}r |
>{\columncolor[HTML]{FFFFFF}}r |
>{\columncolor[HTML]{FFFFFF}}r |}
\hline
\cellcolor[HTML]{FFCB2F}SrTiO\textsubscript{3}                             & \cellcolor[HTML]{FFCB2F}SrTiO\textsubscript{3}                             & \cellcolor[HTML]{FFCB2F}{\color[HTML]{3B2322} LiFePO\textsubscript{4} }     & \cellcolor[HTML]{FFCB2F}Li\textsubscript{2}MnO\textsubscript{3}                            & \multicolumn{1}{r|}{\cellcolor[HTML]{FFCB2F}Sr\textsubscript{3}GaN\textsubscript{3} }      \\ \hline
Sr\_OOO                                                    & \_Ti OOO                                                   & {\color[HTML]{3B2322} Li\_POOOO}                           & Li Li \_ OOO                                              & {\color[HTML]{3B2322} SrSrSr\_NNN}                        \\ \hline
\cellcolor[HTML]{FFFC9E}{\color[HTML]{3B2322} Ti 0.051  } & {\color[HTML]{3B2322} Ba 0.286  }                         & {\color[HTML]{3B2322} Co 0.206  }                         & {\color[HTML]{3B2322} O 0.098 }                          & {\color[HTML]{3B2322} Sr 0.082 }                         \\ \hline
{\color[HTML]{3B2322} Ru 0.046  }                         & {\color[HTML]{3B2322} Ca 0.209  }                         & {\color[HTML]{3B2322} Mn 0.150  }                         & {\color[HTML]{3B2322} Li 0.056 }                         & {\color[HTML]{3B2322} B 0.074 }                          \\ \hline
{\color[HTML]{3B2322} Si 0.044  }                         & \cellcolor[HTML]{FFFC9E}{\color[HTML]{3B2322} Sr 0.144  }                         & \cellcolor[HTML]{FFFC9E}{\color[HTML]{3B2322} Fe 0.129  } & {\color[HTML]{3B2322} C 0.056 }                          & {\color[HTML]{3B2322} Ir 0.046 }                         \\ \hline
{\color[HTML]{3B2322} Ge 0.038  }                         & {\color[HTML]{3B2322} Mg 0.089  }                         & {\color[HTML]{3B2322} Cu 0.120  }                         & {\color[HTML]{3B2322} Si 0.045 }                         & \cellcolor[HTML]{DBF4AD}{\color[HTML]{3B2322} Fe 0.043 } \\ \hline
{\color[HTML]{3B2322} C  0.036  }                         & { Ti 0.045  } & {\color[HTML]{3B2322} Ni 0.117  }                         & \cellcolor[HTML]{FFFC9E}{\color[HTML]{3B2322} Mn 0.040 } & \cellcolor[HTML]{FFFC9E}{\color[HTML]{3B2322} Ga 0.042 } \\ \hline
{\color[HTML]{3B2322} Sn 0.034  }                         & {\color[HTML]{3B2322} Y  0.036  }                         & {\color[HTML]{3B2322} Cr 0.070  }                         & {\color[HTML]{3B2322} Co 0.038 }                         & \cellcolor[HTML]{DBF4AD}{\color[HTML]{3B2322} Cr 0.042 } \\ \hline
{\color[HTML]{3B2322} Mn 0.031  }                         & {\color[HTML]{3B2322} Na 0.032  }                         & {\color[HTML]{3B2322} V  0.033  }                         & {\color[HTML]{3B2322} Cr 0.030 }                         & {\color[HTML]{3B2322} N 0.040 }                          \\ \hline
{\color[HTML]{3B2322} O  0.028  }                         & {\color[HTML]{3B2322} K  0.030  }                         & {\color[HTML]{3B2322} Zn 0.026  }                         & {\color[HTML]{3B2322} Fe 0.029 }                         & {\color[HTML]{3B2322} Co 0.040 }                         \\ \hline
{\color[HTML]{3B2322} Ir 0.028  }                         & {\color[HTML]{3B2322} Li 0.029  }                         & {\color[HTML]{3B2322} Mg 0.023  }                         & {\color[HTML]{3B2322} Ti 0.028 }                         & {\color[HTML]{3B2322} Ge 0.033 }                         \\ \hline
{\color[HTML]{3B2322} Zr 0.027  }                         & {\color[HTML]{3B2322} Rb 0.021  }                         & {\color[HTML]{3B2322} Li 0.021  }                         & {\color[HTML]{3B2322} O 0.028 }                          & {\color[HTML]{3B2322} Mn 0.033 }                         \\ \hline
\end{tabular}
\end{table}

\paragraph{BLMM learns materials chemistry}

To evaluate whether our BLMM model learns the implicit chemical rules for composing feasible materials, we select a dataset of compositions that are charge-neutral, have balanced electronegativity and unique oxidation state assignments as estimated by the Pymatgen oxidation guess module. The last requirement makes it nontrivial to select appropriate elements for substitution. We also require the maximum number of atoms for each element to be less or equal to 10 for fast oxidatiaon states calculation. In total, we obtain 47737 materials compositions. We then expand each of the formulas (e.g. SrTiO$_3$ --> Sr Ti O O O) and randomly mask one element in the sequence and run the blank filling using our BLMM model. We then check the charge neutrality and electronegativity balance after element substitution compared to the performance by random element substitution. Our experiment shows that BLMM can achieve 92.6\% charge neutrality and 90.8\% with balanced electronegativity after BLMM suggested missing element filling. By comparison, the random element substitution can succeed in 89.1\% for charge neutrality and 80.5\% for balanced electronegativity, indicating that our BLMM models have successfully learned the chemical rules of inorganic materials compositions. It should be noted that the surprising random substitution's 80.5\% is due to the replacement of a single atom over a charge-balanced formula with balanced electronegativity.

\FloatBarrier

\subsection{Conditional generative design of materials with high bandgap}

\begin{figure}[ht]
  \centering
  \includegraphics[width=0.6\linewidth]{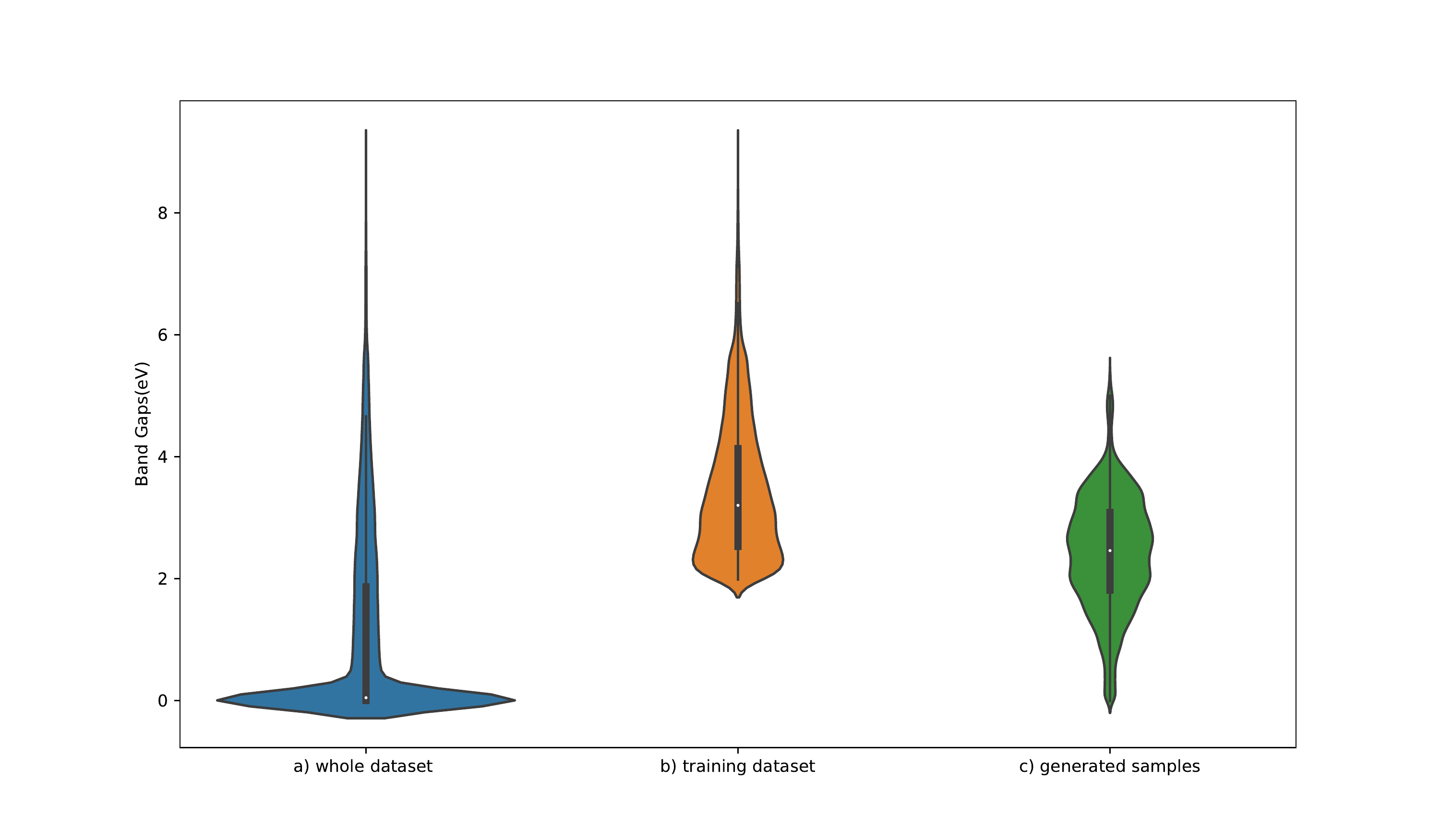}
  \caption{Band-gap distribution for (a) the whole materialsproject materials, (b) the training set of high-band gap materials for BLMM model; (3)  the generated samples. The band gap distribution of the generated ones is much closer to the training set than the whole dataset.}
  \label{fig:bandgap}
\end{figure}

To evaluate whether our language model can capture the composition patterns for high-bandgap materials, we collect 29,772 formulas with band gap above 1.98 eV) from the MaterialsProject (for those formulas with multiple phases, we include it if it has one phase with band gap greater than 2.0 eV). We then train the BLMM language generator and used it to generate 100,000 formulas. We then use the composition based band gap prediction model (See Method) to predict the band gaps of these hypothetical materials and plot their distribution against the band gap distribution of the training set. As shown in Figure \ref{fig:bandgap}, the band gap distribution of our hypothetical compositions is much closer to the training set compared to the band gap distribution of all materials project samples, which indicates that the BLMM-bandgap model has learned the implicit rules to generate high-band gap materials.

\subsection{New materials predicted by our algorithm and validated using DFT}

Due to the difficulty for DFT simulation for compounds with La and Ar family elements, we prepare a subset of ICSD compositions that excludes those elements and use it to train a BLMM model (Detailed hyperparameters are described in Supplementary file). We generate 100,000 ternary and quarternary material compositions and then we predict their formation energy using the composition based formation energy prediction model (See Method). Next, we calculate their total energy and predict their e-above-hull energies to rank these candidates. We then pick the top 100 formulas with the lowest predicted e-above-hull energy and apply our TCSP, a template based crystal structure prediction  algorithm \cite{wei2021tcsp} to obtain the structures. For the predicted structures with the best quality scores, we run DFT relaxation to get the final structures and calculate their formation energy and e-above-hull using DFT method (see Method). 
Table \ref{tab:finding} shows the top 20 discovered new binary and ternary materials along with their formation energy.

\begin{table}[ht]
\centering
\caption{Twenty binary and ternary materials found with negative formation energy ($E_\mathrm{form}$).}
\label{tab:finding}
\begin{tabular}{|c|c|c|c|}
\hline
Formula  & $E_\mathrm{form } (eV)$ & Formula          & $E_\mathrm{form} (eV)$ \\ \hline
RbN$_3$  & -4.7751            & MnAlN$_2$        & -3.7577           \\ \hline
MoN$_2$  & -3.9658            & TaFeN$_2$        & -3.6287           \\ \hline
SrCl$_2$ & -3.4514            & Sr$_3$(NbN2)$_2$ & -3.5682           \\ \hline
LaCl$_3$ & -3.4018            & Sr$_3$MoN$_3$    & -3.4226           \\ \hline
LuCl$_3$ & -3.3315            & Ba$_3$WN$_3$     & -3.4003           \\ \hline
H$_4$N   & -3.3159            & Ba$_7$HfN$_6$    & -3.3053           \\ \hline
RuN$_2$  & -3.3132            & TaCo$_2$N$_3$    & -3.2159           \\ \hline
HfO$_2$  & -3.3044            & La5Si$_3$O$_13$  & -3.2106           \\ \hline
TiF$_4$  & -3.1989            & Sr$_2$BrN        & -3.2100           \\ \hline
TaF$_5$  & -3.1946            & Sr$_4$IrN$_4$    & -3.0930           \\ \hline
\end{tabular}
\end{table}

\section{Discussion}

We developed a transformer based blank filling language model for learning generative design models of materials compositions. The large-scale experiments on both composition generation and  tinkering/element substitution have shown that they have learned strong chemical rules for creating chemically valid compositions or formulas. Especially, compared with previous GAN based generators\cite{dan2020generative}, our probabilistic BLMM model brings much higher explainability of the tinkering suggestions and more control of the generation process of new compositions, which are highly desirable for materials scientists. 

By comparing the distribution of generated samples versus the training samples (Figure\ref{fig:distribution}) of our BLMM model compared to those by MATGAN model (Fig3 of \cite{dan2020generative}), we find our generated samples share much higher composition distribution of known training samples while the GAN based generator tends to create very different composition families indicating that BLMM models are more suitable for exploitation of known chemical space while MATGAN models are  more suitable for exploration of new chemical design space. This major difference may come from their very different modeling and learning mechanisms: BLMM models explicitly learn the chemical contexts dependency (element dependency within known formulas) while the MATGAN models lack such explicit probabilistic modeling components and rely on the neural network models to indirectly approximate their composition to fool the discriminator, leading to less control of following the probabilistic control of the generation process. Another major difference is that our BLMM models are much slower when generating compositions while MATGAN can generate much faster.

For decades, chemists and materials scientists rely on heuristic knowledge or some chemical rules to explore the chemical design space and find new materials. These rules can be described using certain chemical grammars as suggested in \cite{margraf2021heterogeneous}, which can significantly reduce the sampling errors compared to random composition generation while exhaustive enumeration and screening is infeasible since the number of quarternary compounds can already exceed $10^{12}$. While a few heuristic grammar rules can be deduced by human experts, it has the huge risk of missing many unknown grammar rules or some implicit grammars not analytically expressed by the grammars. On this regard, our deep transformer network models have big advantages. First, our model implicitly uses the data-driven strategy to learn the composition generation grammars from the known composition data, which avoids the pitfalls of human-defined chemical grammars. On the one hand our BLMM models have already shown up to 8 times of enrichment when generating charge neutral and electronegativity balanced compositions compared to random generation as shown in Figure \ref{fig:pureChart}. In addition, the probabilistic nature of our model share the certain advantages of the grammar rules in terms of their interpretability. 

While our models can generate chemically valid hypothetical materials compositions in terms of charge neutrality and electronegativity balance, whether these compositions can be synthesized into stable structures remain unknown. While prediction models have been proposed for synthesizability prediction \cite{jang2020structure}, formation energy prediction \cite{omee2021scalable}, and e-above-hull calculation, these models and algorithms usually require the availability of the crystal structures. Fortunately, recent progress in template based \cite{kusaba2022crystal,wei2021tcsp}, deep learning based \cite{hu2021alphacrystal}, and global optimization based crystal structure prediction tools \cite{oganov2012crystal,shao2022symmetry} have made it possible to guess the crystal structures for increasing families of materials, which can be combined with our composition generators to explore and discover new materials.

\section{Materials and Methods}
\label{sec:others}

\subsection{Dataset}

To evaluate the performance of our language model based generator, we trained two sets of models with two different types of datasets. The first category of datasets are all screened formulas from ICSD/MP/OQMD with the number of elements less than 9, number of atoms in unit cell less than 100 and without fractional coordinates. These datasets may contain a certain amount of materials that are not charge neutral or balanced electronegativity. The second category of datasets are those samples from the first category but are charge neutral and have balanced electronegativity. The details of these datasets are shown in Table\ref{tab:datasets}. 

\begin{table}[th]
\centering
\caption{Six datasets used in experiments: pure datasets only include selected samples with neutral charge and  balanced electronegativity; mixed datasets do not have such limits.}
\label{tab:datasets}
\begin{tabular}{|
>{\columncolor[HTML]{FFFFFF}}l 
>{\columncolor[HTML]{FFFFFF}}r 
>{\columncolor[HTML]{FFFFFF}}r 
>{\columncolor[HTML]{FFFFFF}}r |c|
>{\columncolor[HTML]{FFFFFF}}l 
>{\columncolor[HTML]{FFFFFF}}r 
>{\columncolor[HTML]{FFFFFF}}r 
>{\columncolor[HTML]{FFFFFF}}r |}
\hline
\multicolumn{4}{|c|}{\cellcolor[HTML]{FFD966}Mix datasets}                                                                                                                                                        & \multicolumn{1}{l|}{ \cellcolor[HTML]{FFD966} }                      & \multicolumn{4}{c|}{\cellcolor[HTML]{FFD966}Pure datasets}                                                                                                                                                       \\ \hline
\multicolumn{1}{|l|}{\cellcolor[HTML]{FFFFFF}}      & \multicolumn{1}{l|}{\cellcolor[HTML]{FFFFFF}ICSD-mix}  & \multicolumn{1}{l|}{\cellcolor[HTML]{FFFFFF}OQMD-mix}   & \multicolumn{1}{l|}{\cellcolor[HTML]{FFFFFF}MP-mix} & \cellcolor[HTML]{FFFFFF}                   & \multicolumn{1}{l|}{\cellcolor[HTML]{FFFFFF}}      & \multicolumn{1}{l|}{\cellcolor[HTML]{FFFFFF}ICSD-pure}  & \multicolumn{1}{l|}{\cellcolor[HTML]{FFFFFF}OQMD-pure}   & \multicolumn{1}{l|}{\cellcolor[HTML]{FFFFFF}MP-pure} \\ \cline{1-4} \cline{6-9} 
\multicolumn{1}{|l|}{\cellcolor[HTML]{FFFFFF}Total} & \multicolumn{1}{r|}{\cellcolor[HTML]{FFFFFF}52317} & \multicolumn{1}{r|}{\cellcolor[HTML]{FFFFFF}363182} & 89121                                           & \cellcolor[HTML]{FFFFFF}                   & \multicolumn{1}{l|}{\cellcolor[HTML]{FFFFFF}Total} & \multicolumn{1}{r|}{\cellcolor[HTML]{FFFFFF}39431} & \multicolumn{1}{r|}{\cellcolor[HTML]{FFFFFF}216540} & 63703                                           \\ \cline{1-4} \cline{6-9} 
\multicolumn{1}{|l|}{\cellcolor[HTML]{FFFFFF}Train} & \multicolumn{1}{r|}{\cellcolor[HTML]{FFFFFF}50755} & \multicolumn{1}{r|}{\cellcolor[HTML]{FFFFFF}345022} & 84664                                           & \cellcolor[HTML]{FFFFFF}                   & \multicolumn{1}{l|}{\cellcolor[HTML]{FFFFFF}Train} & \multicolumn{1}{r|}{\cellcolor[HTML]{FFFFFF}37459} & \multicolumn{1}{r|}{\cellcolor[HTML]{FFFFFF}205713} & 60517                                           \\ \cline{1-4} \cline{6-9} 
\multicolumn{1}{|l|}{\cellcolor[HTML]{FFFFFF}Valid} & \multicolumn{1}{r|}{\cellcolor[HTML]{FFFFFF}1336}  & \multicolumn{1}{r|}{\cellcolor[HTML]{FFFFFF}9080}   & 2228                                            & \cellcolor[HTML]{FFFFFF}                   & \multicolumn{1}{l|}{\cellcolor[HTML]{FFFFFF}Valid} & \multicolumn{1}{r|}{\cellcolor[HTML]{FFFFFF}986}   & \multicolumn{1}{r|}{\cellcolor[HTML]{FFFFFF}5413}   & 1593                                            \\ \cline{1-4} \cline{6-9} 
\multicolumn{1}{|l|}{\cellcolor[HTML]{FFFFFF}Test}  & \multicolumn{1}{r|}{\cellcolor[HTML]{FFFFFF}1336}  & \multicolumn{1}{r|}{\cellcolor[HTML]{FFFFFF}9080}   & 2228                                            & \multirow{-5}{*}{\cellcolor[HTML]{FFFFFF}} & \multicolumn{1}{l|}{\cellcolor[HTML]{FFFFFF}Test}  & \multicolumn{1}{r|}{\cellcolor[HTML]{FFFFFF}986}   & \multicolumn{1}{r|}{\cellcolor[HTML]{FFFFFF}5413}   & 1593                                            \\ \hline
\end{tabular}
\end{table}

\subsection{Baseline pseudo random composition generator:}
We create a pseudo random composition generator as the baseline. For all generated samples, we count the numbers of samples with different number of elements from 2 to $E_n$. Then for each of the element number $K$, we generate the same number of composition samples with $K$ elements. For each of them, we randomly pick an atom number from 1 to 20 for each of the $K$ elements. This will ensure the distribution of binary, ternary and etc. samples are the same as the comparison group.

\subsection{DFT calculations}

To check the structure stability of the predicted materials, we apply the first-principles calculations based on the density functional theory (DFT) using the Vienna \textit{ab initio} simulation package (VASP) \cite{Vasp1,Vasp2,Vasp3,Vasp4}. The projected augmented wave (PAW) pseudopotentials, where 520 eV plane-wave cutoff energy, are used to treat the electron-ion interactions \cite{PAW1, PAW2}. The exchange-correlation functional is considered with the generalized gradient approximation (GGA) based on the Perdew-Burke-Ernzerhof (PBE) method \cite{GGA1, GGA2}. The energy convergence criterion is set as 10$^{-5}$ eV, while the atomic positions are optimized with the force convergence criterion of 10$^{-2}$ eV/{\AA}. The Brillouin zone integration for the unit cells was computed using the $\Gamma$-centered  Monkhorst-Pack $k$-meshes. The Formation energies (in eV/atom) of several materials are determined based on the expression in  Eq.~\ref{eq:form}, where $E[\mathrm{Material}]$ is the total energy per unit formula of the considered structure, $E[\textrm{A}_i]$ is the energy of $i^\mathrm{th}$ element of the material, $x_i$ indicates the number of A$_i$ atoms in a unit formula, and $n$ is the total number of atoms in a unit formula($n=\sum_i x_i$).

\begin{equation}
    E_{\mathrm{form}} =\frac{1}{n}(E[\mathrm{Material}] - \sum_i x_i E[\textrm{A}_i])
    \label{eq:form}
\end{equation}

\subsection{Evaluation criteria}

The performance of materials generative models can be mainly evaluated using three criteria including validity, uniqueness and recovery rate \cite{dan2020generative}. Here the validity of generated samples are evaluated using the charge neutrality and electronegativity balance, which are two fundamental chemical rules of crystals. It is interesting to check how the generated samples from our BLMM models satisfy these rules without explicit enforcement of such rules during model training. To do this, we adopt the charge-neutrality and electronegativity check procedure as proposed in ref \cite{davies2019smact} to calculate the percentages of samples that obey these rules within the training and generated sets. 

The uniqueness of a generative model measures the percentage of the number of unique samples out of the number of all generated samples (n). The higher this measure, the better capability the model can generate diverse samples. 

The recovery rate measures the percentage of samples from the training or testing set that have been re-generated by the generator model. The high recovery rate over the test set indicates that a generator has high discovery performance since the test set samples are known crystals that actually exist. A related measure is the novelty of a generator, which measures the percentage of the generated samples are new samples that do not exist before.

\subsection{Hyper-parameters}

We conduct hyper-parameter studies of our BLMM model using the ICSD-pure dataset to evaluate how the major parameters affect our model performance. We train a set of BLMM models with a maximum number of epochs of 3000 and different hyper-parameter configurations. To evaluate their composition generation performance, we use each of these models to generate 10,000 formulas and calculate the percentages of charge-neutral (CN) and electronegativity-balanced (EN) samples out of all the generated samples. The hyper-parameters evaluated here include: the number of transformer layers, the number of transformer heads, the size of the hidden layers, dropout rate and learning rate. Instead of exhaustive enumeration of all possible parameter configurations, we use a default hyper-parameter set which includes six transformer layers, 8 transformer heads and the size of the hidden layer is 2048. The default dropout rate and learning rate are 0.3 and 0.0001. Then each time, we change one hyper-parameter and calculate their CN/EN percentages. In addition, we also compare the performance of two other related language models LBLM and INST\cite{shen2020blank} using the same default hyper-parameter set. 

\begin{figure}[ht!] 
    \begin{subfigure}[t]{0.5\textwidth}
        \includegraphics[width=\textwidth]{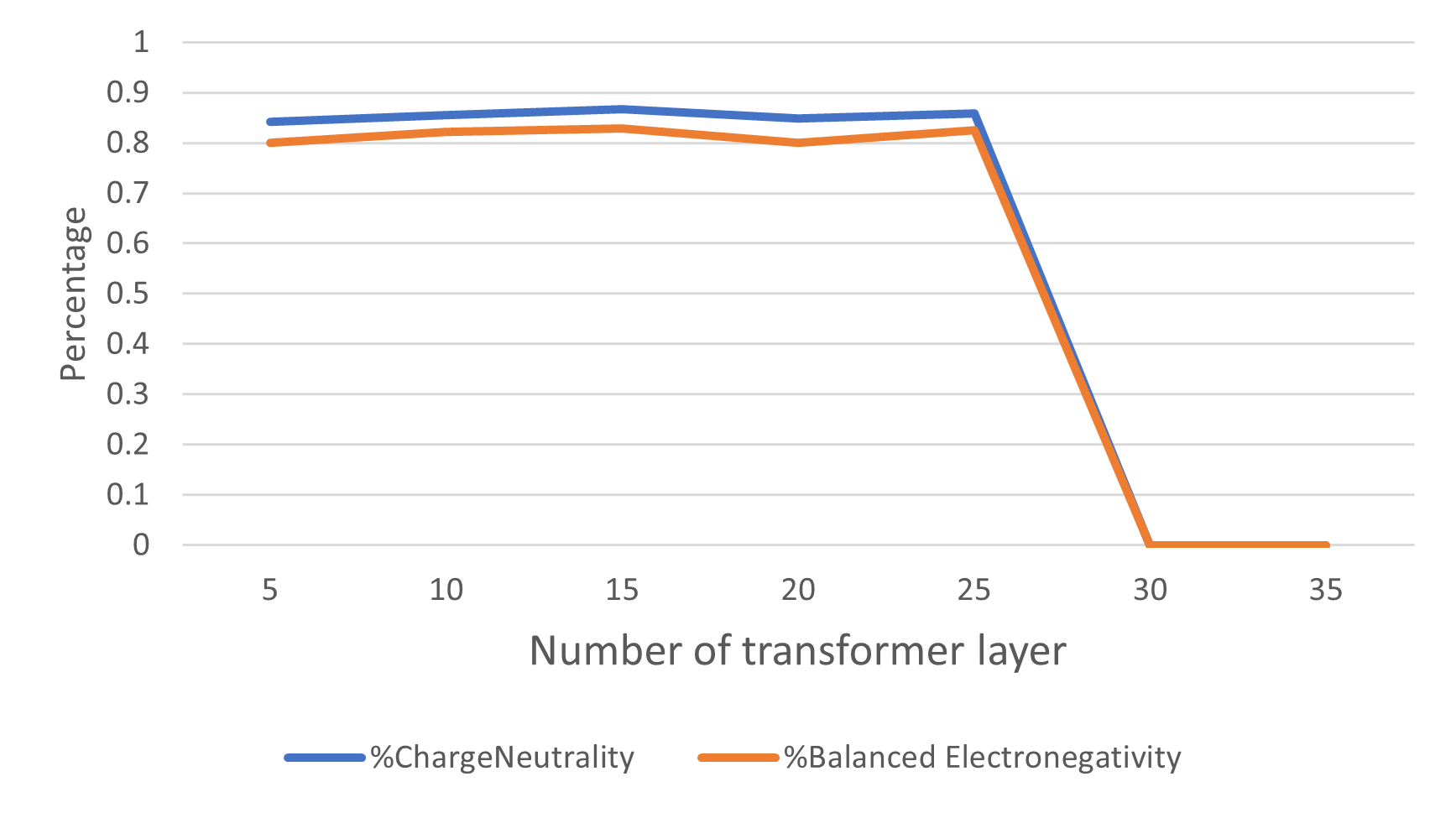}
        \caption{}
        \vspace{-3pt}
        \label{fig:hyper_layers}
    \end{subfigure}
    \begin{subfigure}[t]{0.5\textwidth}
        \includegraphics[width=\textwidth]{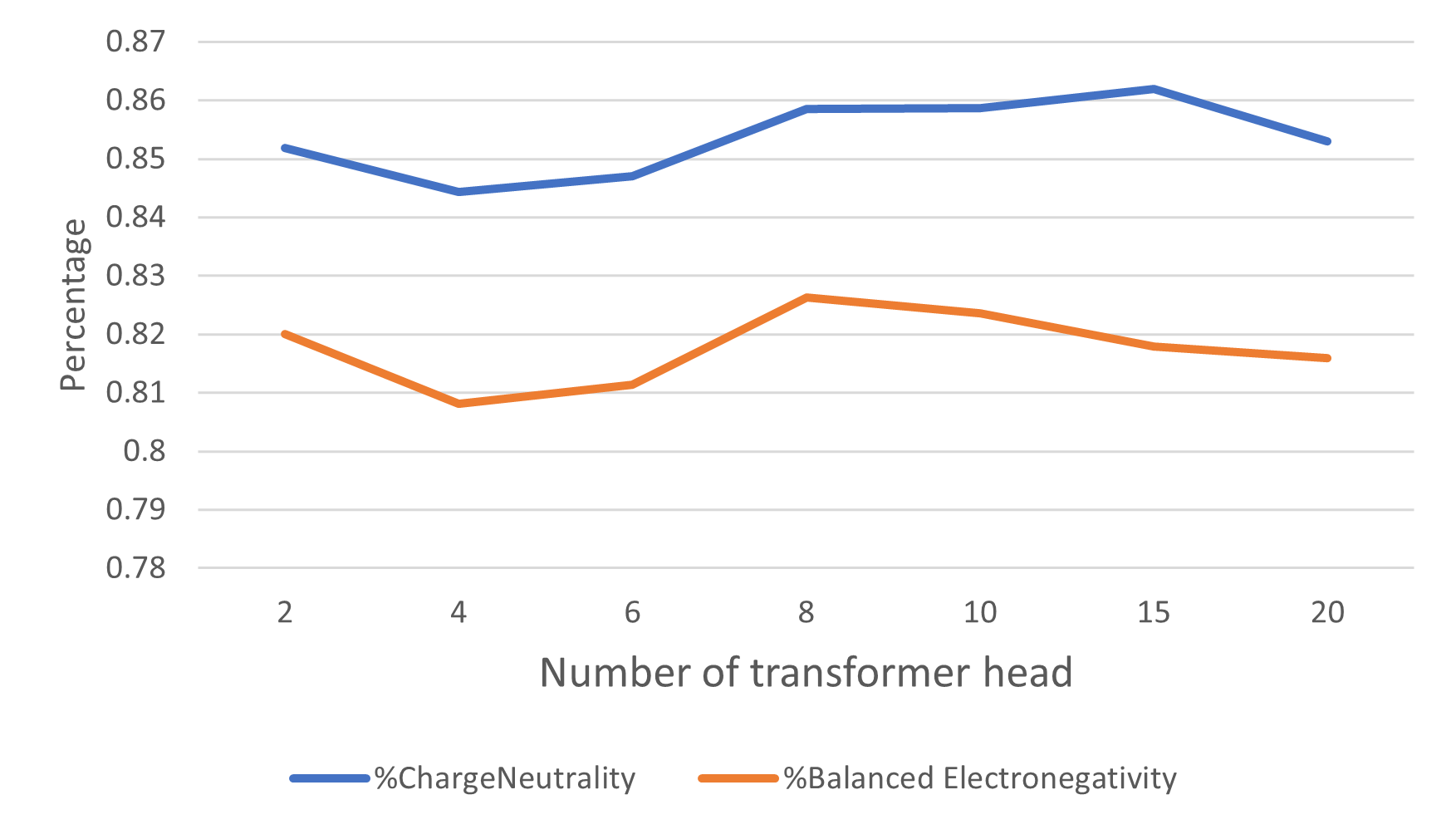}
        \caption{}
        \vspace{-3pt}
        \label{fig:hyper_hid}
    \end{subfigure} 
 \begin{subfigure}[t]{0.50\textwidth}
        \includegraphics[width=\textwidth]{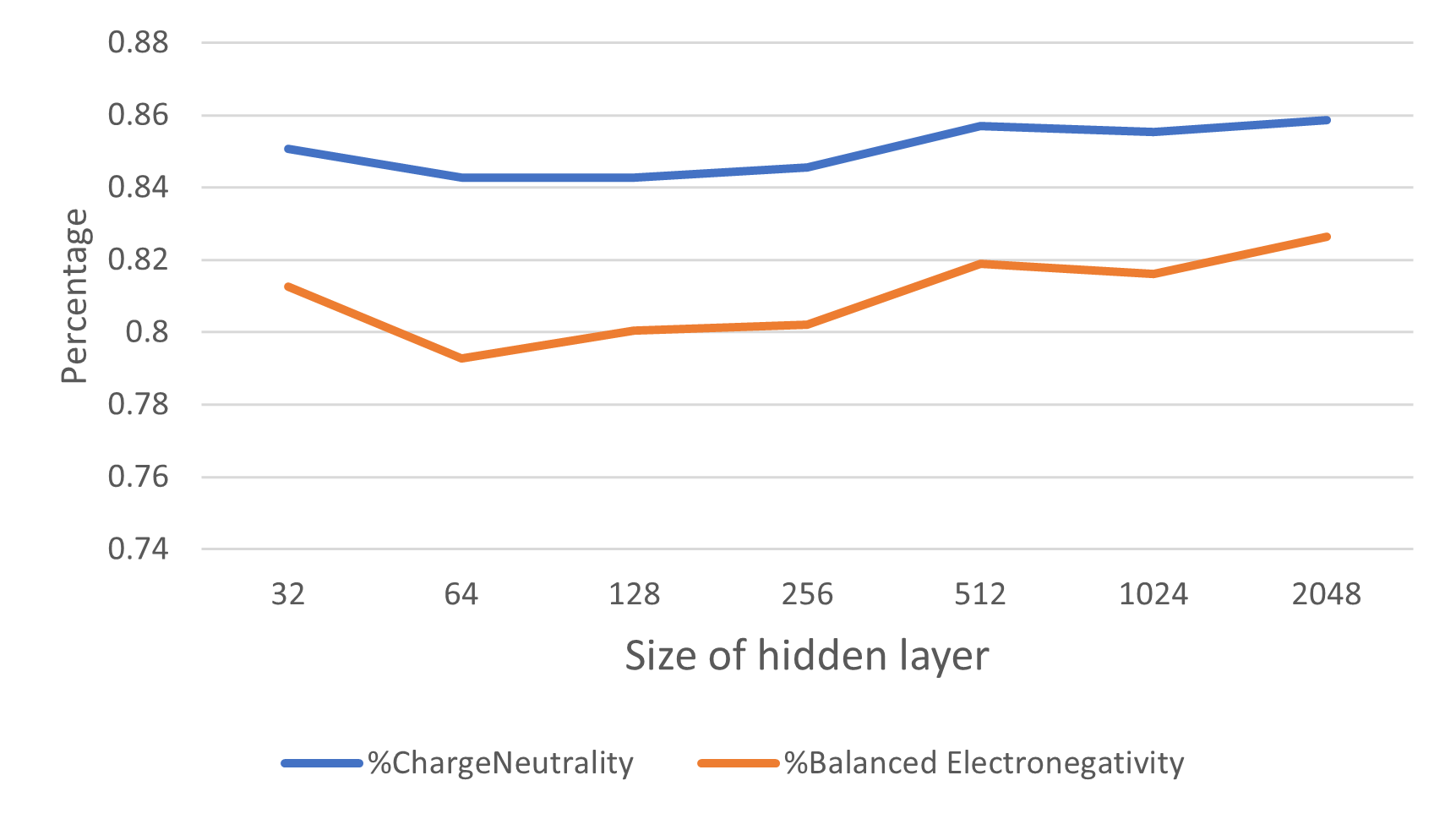}
        \caption{}
        \vspace{-3pt}
        \label{fig:hyper_head}
    \end{subfigure}              
    \begin{subfigure}[t]{0.50\textwidth}
        \includegraphics[width=\textwidth]{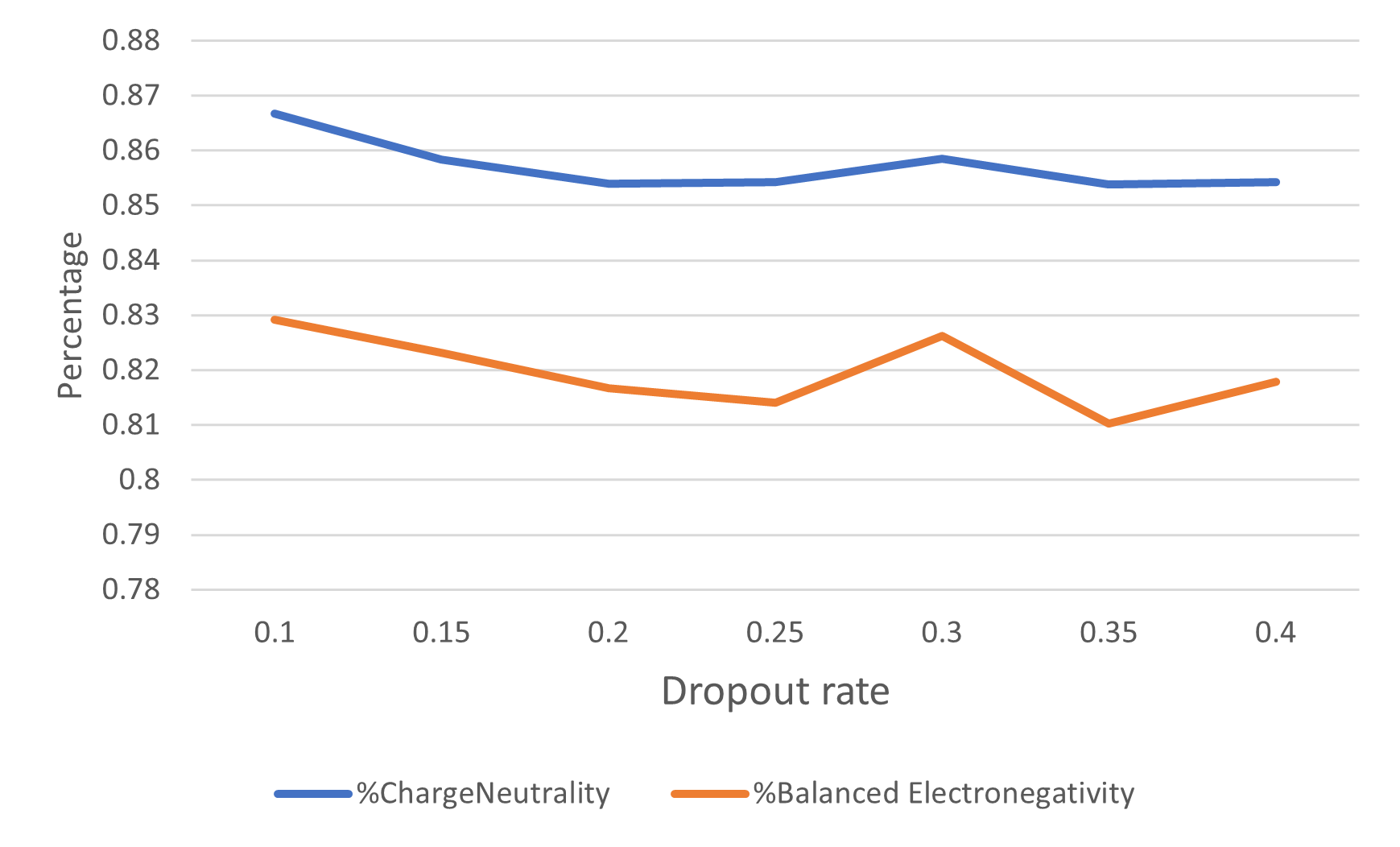}
        \caption{}
        \vspace{-3pt}
        \label{fig:hyper_dropout}
    \end{subfigure}     
    \begin{subfigure}[t]{0.5\textwidth}
        \includegraphics[width=\textwidth]{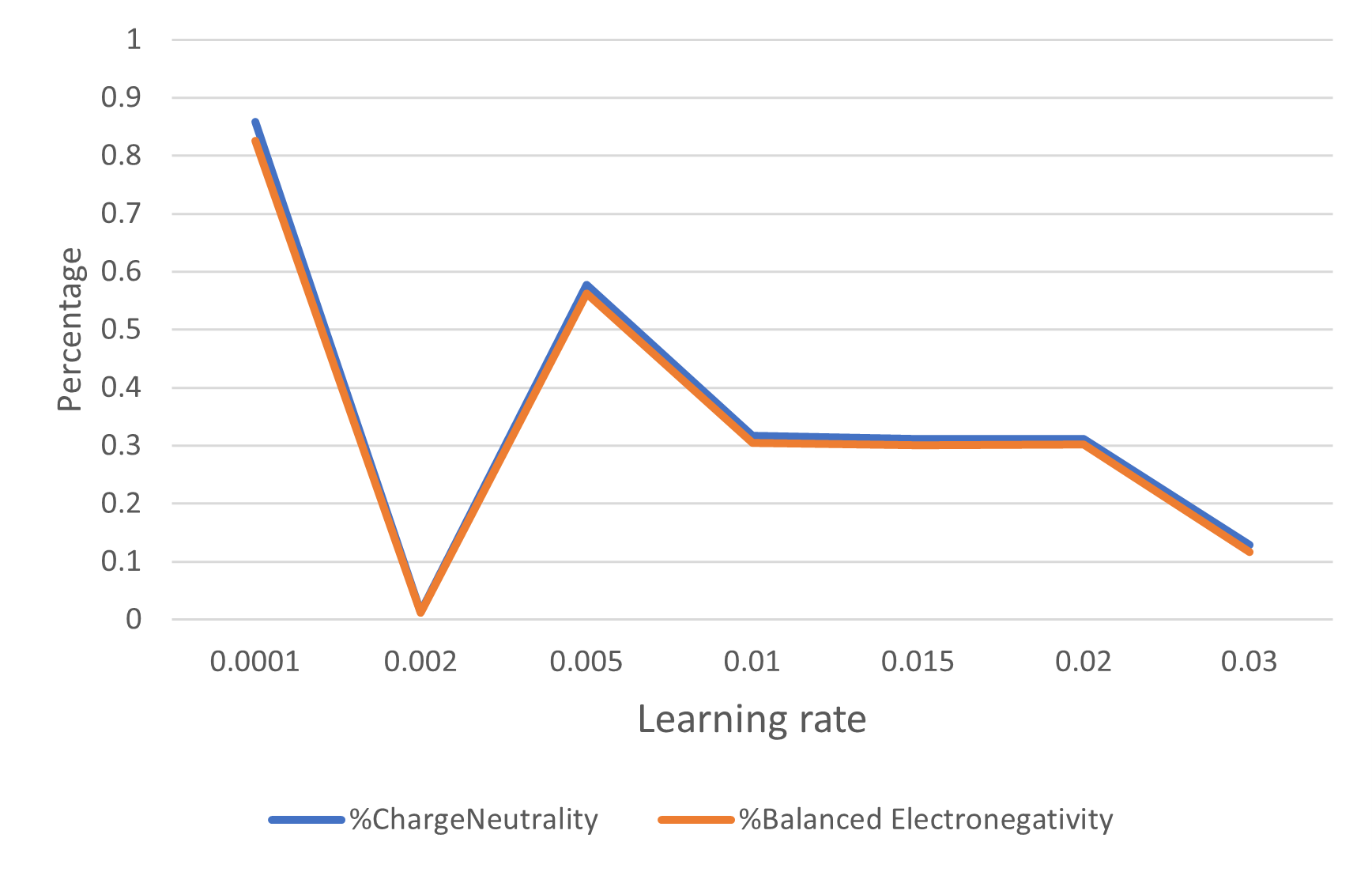}
        \caption{}
        \vspace{-3pt}
        \label{fig:hpyer_lr}
    \end{subfigure}
     \begin{subfigure}[t]{0.50\textwidth}
        \includegraphics[width=\textwidth]{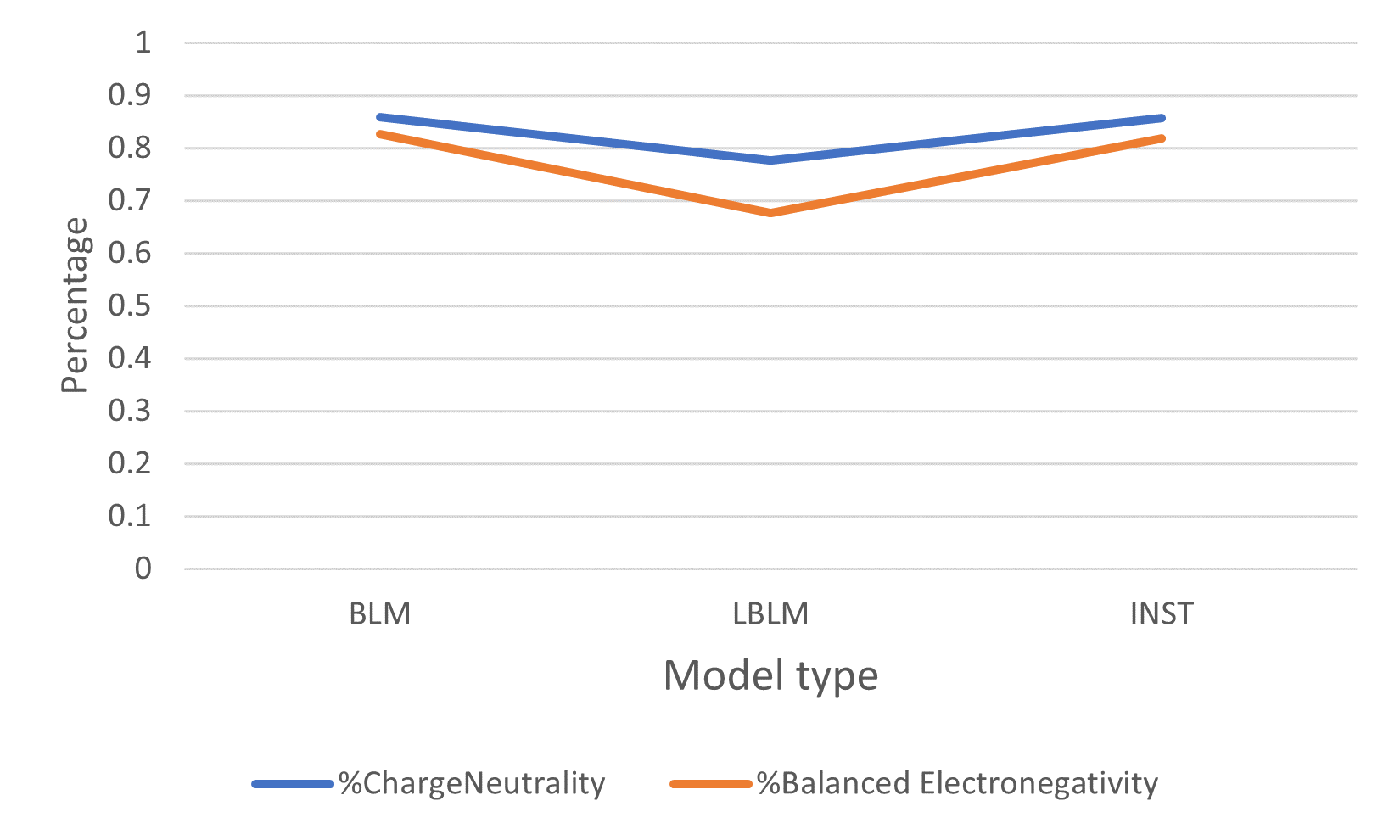}
        \caption{}
        \vspace{-3pt}
        \label{fig:hyper_inst_lblm}
    \end{subfigure}              
   \caption{Hyper-parameter tuning of BLMM materials composition generator. (a) The percentages of charge-neutral (CN) and electronegativity-balanced (EN) samples out of all generated samples by the BLMM models trained with different number of the transformer layers (b) The CN/EN percentages of the models trained with different number of the transformer heads (c) The CN/EN percentages of the models trained with different sizes of the hidden layer (d) The CN/EN percentages of the BLMM models trained with different dropout rate (e) The CN/EN percentages of the BLMM models trained with different learning rates (f) The CN/EN percentages of the BLMM models compared to LBLM model and INST model.}
  \label{fig:hyperparameter}
\end{figure}

First, we check how the number of transformer layers affects the CN/EN percentages of generated samples. As shown in Figure \ref{fig:hyper_layers}, these trained models achieve good performance when the number of transformer layers range from 5 to 25 without significant difference. The models achieve the charge-neutrality (CN) percentages of 84.26\%, 85.56\%,  86.64\%, 0.8494\%, 85.91\% and EN percentages of 79.97\%, 82.11\%, 82.88\%, 80.02\%, 82.48\% for for 5/10/15/20/25 transformer layers respectively. However, when the number of transformer layers reaches 30 and 35, these models could not generate a contiguous sequence of elements but insert a $<blank>$ between every two elements in the generated sequence, leading to invalid formulas.

Second, we evaluate how the number of transformer heads affects the BLMM model performance. Figure \ref{fig:hyper_head} shows that the model has a good performance in terms of CN/EN percentages when the number of transformer heads is 8. When the number of transformer heads reaches 15, the model has the best CN percentage of 86.19\%, which is though not significant from the performance of the model with 8 transformer heads.

Third, we run experiments with different sizes of the inner hidden layers ranging from 32 to 2048. We vary the hidden layer sizes from 32 to 2048 (the default size of the hidden layer). Figure \ref{fig:hyper_hid} shows that the models with larger sizes of hidden layers (512/1024/2048) achieve better performance compared to the models with smaller hidden layer sizes. We further evaluate the model performance changes with different dropout rates ranging from 0.1 to 0.4. Figure \ref{fig:hyper_dropout}, the models have good performance when the dropout rate are 0.1 and 0.3. The impact of the learning rate over generation performance is shown in Figure  \ref{fig:hpyer_lr}. We find that the model performs well for with CN/EN percentages of 85.85\% and 82.63\% respectively only when the learning rate is 0.0001 (the default learning rate). We determine the maximum number of epochs to run for our training by plotting the training and validation loss curves.
We found that for most of our model training, the losses converge within 3000 epochs. So we have chosen 3000 as the no. of epochs to train our models. 

Finally, we compare the performance of BLMM with two other related language models: LBLM and INST \cite{shen2020blank}. As shown in Figure \ref{fig:hyper_inst_lblm}, BLMM achieves the best performance in both charge-neutrality and balanced electronegativity (EN) percentages, with 85.85\% and 82.63\% respectively. The INST model also has good performance with CN/EN percentages of 85.69\% and 81.76\% respectively.

\subsection{Formation energy and bandgap prediction models based on Roost}

To check the quality of generated compositions, we train composition based prediction models for both formation energy and band gaps using the dataset downloaded from materialsproject database \cite{jain2013commentary}. The machine learning model we used is roost, a graph message passing neural network as described
in \cite{goodall2020predicting}. The training set of Roost-FE contains 125,613 unique compositions. All compositions with multiple phases will only keep the lowest formation energy records. The Roost-Bandgap model is trained with 113,501 samples. The formation energy roost model achieves an MAE of 70.181 eV while the band gap predictor achieves an MAE of 0.6645 eV as evaluated on the 10\% hold-out test sets.

\FloatBarrier


\bibliographystyle{unsrt}  
\bibliography{references}  

\section*{Acknowledgement}

\paragraph{Funding:}The research reported in this work was supported in part by National Science Foundation under the grant and 1940099 and 1905775. The views, perspectives, and content do not necessarily represent the official views of the NSF. 
\paragraph{Author contributions:}
Conceptualization, J.H.; methodology,J.H. L.W., Q.L, D.S., Y.S.; software, J.H.,L.W.,S.S.,Y.S.; resources, J.H.; writing--original draft preparation, J.H., L.W., E.S.; writing--review and editing, J.H, L.W., D.S., F.C.; visualization, L.W., Y.S., Q.L, J.H.; supervision, J.H.;  funding acquisition, J.H.
\paragraph{Competing interests: The authors declare that they have no competing interests}

\end{document}